\documentclass[aps,pra,nobibnotes,twocolumn,superscriptaddress]{revtex4-1}
\usepackage{graphicx}
\usepackage{mathrsfs}
\usepackage{bm}
\usepackage{amsmath}
\usepackage{dcolumn}
\usepackage{epstopdf}
\usepackage{dsfont}
\usepackage{amssymb}
\usepackage{tabularx}
\usepackage{array}
\usepackage{float}
\usepackage{color}
\usepackage{epstopdf}
\usepackage{mathrsfs}
\usepackage[colorlinks, linkcolor=blue,anchorcolor=blue,citecolor=blue,urlcolor=blue]{hyperref}
\emergencystretch=\maxdimen
\begin{document}

\preprint{APS/123-QED}

\title{Self-Organized Bosonic Domain Walls}

\author{Xingchuan Zhu}
\email{These authors contributed equally}
\affiliation{Department of Physics, Beijing Normal University, Beijing,
100875, China}
\affiliation{Department of Physics, Key Laboratory of Micro-Nano
Measurement-Manipulation and Physics (Ministry of Education), Beihang
University, Beijing, 100191, China}

\author{Shiying Dong}
\email{These authors contributed equally}
\affiliation{Department of Physics, Key Laboratory of Micro-Nano
Measurement-Manipulation and Physics (Ministry of Education), Beihang
University, Beijing, 100191, China}

\author{Yang Lin}
\affiliation{Department of Physics, Key Laboratory of Micro-Nano
Measurement-Manipulation and Physics (Ministry of Education), Beihang
University, Beijing, 100191, China}

\author{Rubem Mondaini}
\email{rmondaini@csrc.ac.cn}
\affiliation{Beijing Computational Science Research Center, Beijing
100084, China}

\author{Huaiming Guo}
\email{hmguo@buaa.edu.cn}
\affiliation{Department of Physics, Key Laboratory of Micro-Nano
Measurement-Manipulation and Physics (Ministry of Education), Beihang
University, Beijing, 100191, China}

\author{Shiping Feng}
\affiliation{Department of Physics, Beijing Normal University, Beijing,
100875, China}

\author{Richard T. Scalettar}
\affiliation{Physics Department, University of California, Davis, CA
95616, USA}


\begin{abstract}
 Hardcore bosons on honeycomb lattice ribbons with zigzag edges are
studied using exact numerical simulations. We map out the phase diagrams
of ribbons with different widths, which contain superfluid and
insulator phases at various fillings. We show that
charge domain walls are energetically favorable, in sharp contrast to
the more typical occupation
of a set of sites on a single sublattice of the bipartite geometry at
$\rho=\frac{1}{2}$ filling. This `self-organized domain wall' separates
two charge-density-wave (CDW) regions with opposite Berry curvatures.
Associated with the change of topological properties, superfluid
transport occurs down the domain wall. Our results provide a concrete context to
observe bosonic topological phenomena and can be simulated
experimentally using bosonic cold atoms trapped in designed optical
lattices.
\end{abstract}

\maketitle

\textit{Introduction.-}
One of the most interesting properties of condensed matter systems is
their condensation into ordered low temperature phases,
breaking an underlying symmetry of the Hamiltonian.  Such phases
typically minimize the free energy $F$; coexistence of the distinct ordered
patterns involves a domain wall, increasing $F$.  Nevertheless,
domain walls often exist in practice in experiments (or in
simulations) as a consequence of long annealing times.
This is especially the case in the presence of disorder which
can pin their motion.

In addition to being manifest as meta-stable states, domain walls
can also arise in other ways.  An important example is provided by
doping away from the commensurate antiferromagnetic (AF) filling of the
cuprate superconductors\cite{tranquada13}, or the Hubbard and $t$-$J$
models that describe them\cite{zaanen89,white03,hellberg99}.
Dopants do not spread uniformly, but instead form ``charge stripes''.
Across these stripes there is a `$\pi$-phase shift' of the AF
order\cite{fradkin15}.
The up-spin occupied sublattice interchanges across
the stripe, realizing a domain wall.

In model Hamiltonian studies on `ladder' geometries
using the density matrix renormalization group,
the charge patterns are found to be
`vertical stripes', i.e.~the doped
holes lie parallel to the {\it short} direction of the cluster\cite{white03}.  These
charge patterns are fundamentally connected not only to magnetism, e.g.
the $\pi$ phase shift, but also to charge density wave and $d$-wave
pairing order.  Studies of stripe physics and the associated domain
walls remain of great interest\cite{zheng17,huang17}, with the possible
coexistence of
Luther-Emery liquid states in which the spin excitations are gapped, and
quasi-long range superconducting correlations
being a key issue \cite{zheng18}.

\begin{figure}[htbp]
\centering \includegraphics[width=8.5cm]{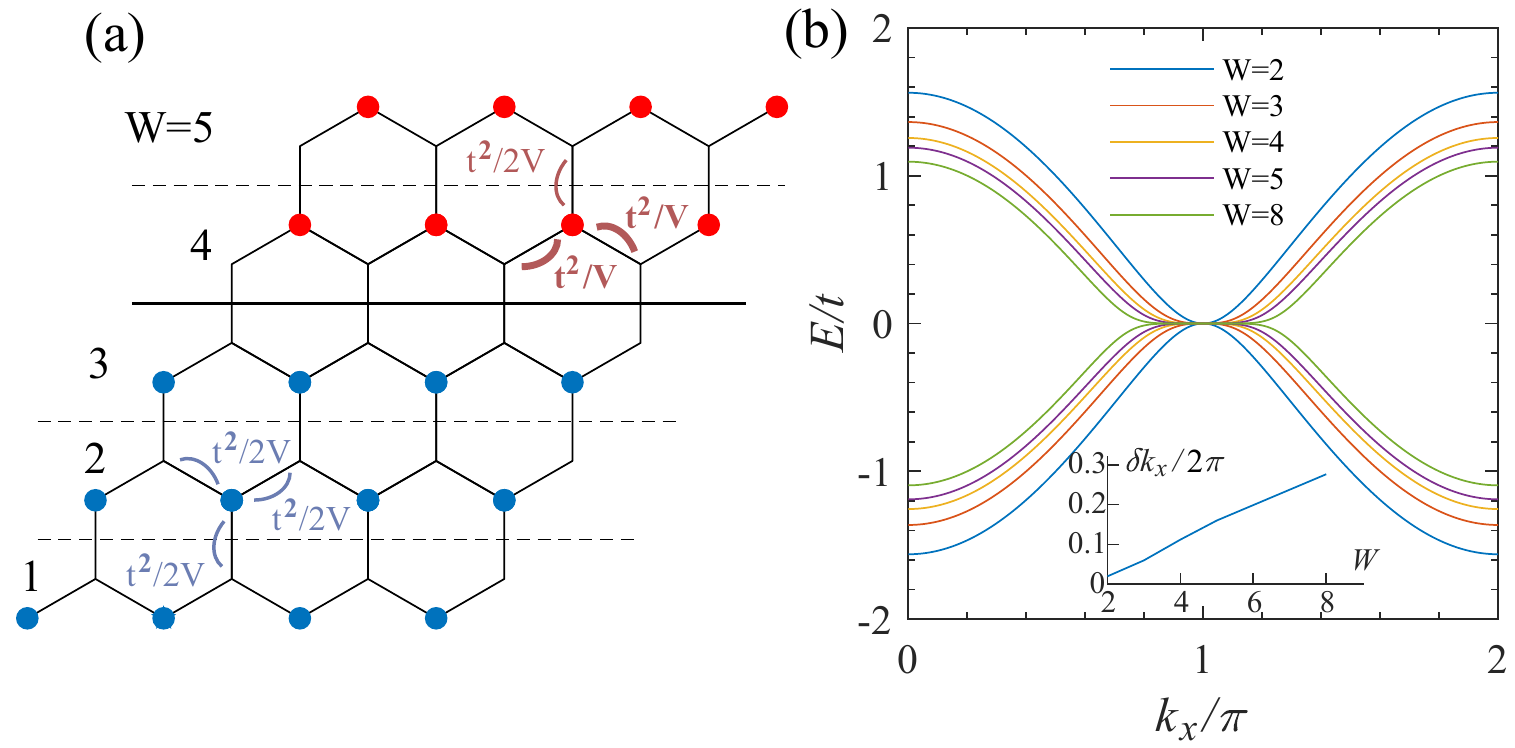} \caption{(a):
Schematics of a width $W=5$ and length $L=4$ honeycomb lattice ribbon with zigzag
edges. One of the (atomic limit) degenerate configurations of the $\rho=\frac{1}{2}$
insulator is shown. The filled circles represent
hardcore bosons, with the colors distinguishing the sublattices. The
domain wall is marked by the solid line. The dashed lines mark the
domain walls of other degenerate configurations. The kinetic energy
gains by second-order hopping processes are marked.
(b): Low-energy bands of ribbons with several $W$
and $L=\infty$.
Inset shows the ratio of the range $\delta k_x$
of the flat band to the full one-dimensional Brillouin zone (BZ),
as a function of $W$.
$\delta k_x/2\pi \sim 1/3$ for large $W$.}
\label{fig1}
\end{figure}

In this Letter we study bosonic particles on honeycomb ribbons.
We discuss four novel features of this geometry.  First, we
argue that charge domain walls are energetically {\it favorable}
compared to occupation of a set of sites on a single sublattice of the
bipartite geometry, even at half-filling.
This is a rather unique feature compared to
situations in which domain walls are excitations rather than the
ground state.  Second, the low density sites of the domain wall are
arranged `horizontally' (parallel to the long axis), rather than
vertically.  These `self-organized domain walls' open the possibility of
superfluid transport down the chain.
Third, associated with this physics
is a non-trivial Berry curvature,
which changes sign across the domain wall.
Finally, the system realizes an exotic one-dimensional
supersolid.

\textit{The model and method.-}
We consider spinless hardcore bosons on zigzag ribbons of a honeycomb lattice,
described by the extended Bose-Hubbard model~\cite{bloch2008, Cazalilla2011}
\begin{eqnarray}\label{eq1}
H=-t\sum_{\langle i,j\rangle}
(\, b_i^{\dagger}b^{\phantom{\dagger}}_{j}+\text{H.c.})+V\sum_{\langle i,j \rangle} n_i n_j-\mu \sum_i n_i.
\end{eqnarray}
$b^{\phantom{\dagger}}_i(b_i^{\dagger}$) are hardcore
boson annihilation(creation) operators,
$n_i=b_i^{\dagger}b^{\phantom{\dagger}}_i$ is the number operator.
Hardcore bosons obey commutation relations
$[b^{\phantom{\dagger}}_i,b_j^{\dagger}]=0$ for sites $i\neq j$ and on-site
anticommutation relations $\{b^{\phantom{\dagger}}_i,b_i^{\dagger}\}=1$.
The first term in
Eq.~(\ref{eq1}) describes nearest-neighbor (NN) hopping, with
amplitude $t$ taken as the unit of energy $(t=1)$.
The second term in Eq.~(\ref{eq1}) is the NN interaction
$V$. Finally, $\mu$ denotes the chemical
potential, which controls the number of bosons in the system.
The model in Eq.~(\ref{eq1}) has a $U(1)$ symmetry, and is
invariant under the transformation $b_j\rightarrow e^{{\rm i}\theta}b_j$ where
$\theta$ is a real-valued phase. This symmetry is spontaneously broken
in a superfluid phase.
The Hamiltonian is invariant under the inversion transformation center
of honeycomb lattice ribbons with zigzag edges.

The band structure of the honeycomb lattice consists of two inequivalent
Dirac points, which are characterized by $\pm \pi$ Berry
phases\cite{castro2009}. As a result of the nontrivial topological
property, localized flat bands connecting the two Dirac points appear on the
zigzag edges\cite{fujita1996,nakada1996,louie2006,yao2009}.
Figure \ref{fig1}(b) shows the low-energy bands.
As the widths of the ribbons are reduced, the lengths of the
flat bands are shortened. The band bottom corresponds to the chemical
potential at which the hardcore bosons begin to fill into the system,
which determines the phase boundary for $\rho=0$.

We employ the stochastic series
expansion (SSE) quantum Monte Carlo (QMC) method~\cite{Bauer2011} with
directed loop updates to study Eq.(1). SSE
expands the partition function in a power series and the trace is written
as a sum of diagonal matrix elements. The directed loop updates
and the fact that the discrete configuration space can be sampled
without floating point operations make the
approach very efficient~\cite{sandvik2002,fabien2005,pollet2004}. Our
simulations are on finite lattices with the total number of sites
$N=2\times W \times L$ with $W$ the width and $L$ the length of a
ribbon[see Fig.\ref{fig1}(a)].
The temperature is set to be low
enough to obtain the ground-state properties.
We also use the exact
diagonalization (ED) method, which is numerically exact, but has strong size
limitations.

\begin{figure}[htbp]
\centering \includegraphics[width=9.cm]{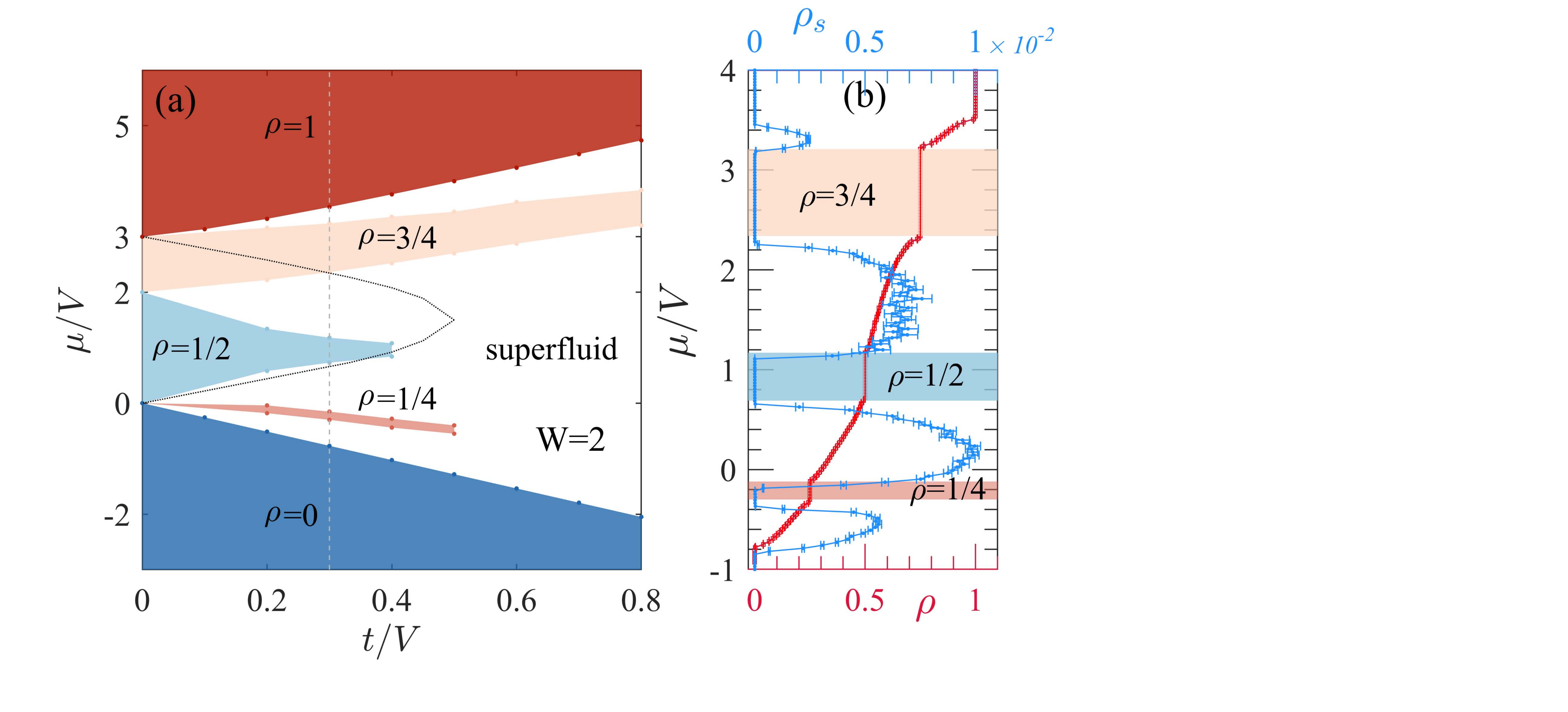} \caption{
(a) Phase diagrams of $W=2$ honeycomb lattice ribbons
contain superfluid regions and insulator phases
at specific fillings.
Dotted lines enclose the $\rho=\frac{1}{2}$ CDW
insulator of hardcore bosons on a periodic two-dimensional honeycomb
lattice\cite{Wessel2007,Nakafuji2017,gan2007}. (b) The average
density and superfluid density as a function of $\mu$ at $t/V=0.3$, which is marked by the dashed line in (a). Ribbon
length $L=24$.}
\label{fig2}
\end{figure}

\textit{The phase diagram.-}
Phase diagrams of ribbons with $W=2$ are shown in
Fig.~\ref{fig2}(a) (see others in Ref.[28]).
In the atomic limit ($t/V=0$),
the density abruptly jumps from empty ($\rho=0$) to half-filled
($\rho=\frac{1}{2}$)
at $\mu/V=0$, with bosons placed so they never occupy adjacent
sites.
At $\rho=\frac{1}{2}$, no further
bosons can be added without being neighbors,
costing energy $\propto V$:  there is a jump in $\mu$.
If the half-filled bosons are placed so that they
occupy only a single sublattice,
the empty sites of one of the boundaries
are special:  they interact with only two neighboring occupied sites.
Thus the $\rho=\frac{1}{2}$ CDW insulator terminates at $\mu/V=2$\cite{cdw}.
Once these special sites are completely occupied, the increase in
density pauses again until $\mu/V=3$, at which point bosons are
added to the remaining
empty sites with three occupied neighbors,
completely filling the lattice.
This atomic limit picture explains the positions of the
insulating lobes bases in Fig.~\ref{fig2}(a).

The $\rho=\frac{1}{2}$ insulator for $0<\mu/V<2$ has a $(W+1)$ fold
degeneracy, including two single sublattice CDWs and $(W-1)$
configurations with a domain wall arranged along each row of vertical
bonds.
The key observation for the presence of self-organized domain walls of
Fig.~\ref{fig2}(a) is that
a domain wall has a {\it pair} of `edges' where empty sites have
only two occupied neighbors.
This greater multiplicity
of special sites compared to a single sublattice
leads to a lower energy at
finite hopping $t/V \neq 0$: the second order energy
decrease when a boson hops
onto a special site is $-t^2/V$ compared to the
$-t^2/2V$ for hopping within the CDW.
Fig.~\ref{fig1}(a) illustrates these different hopping
processes, and the larger overall energy decrease,
$-5t^2/2V$, of a site adjacent to a domain wall
compared to $-3t^2/2V$ gained by a boson inside the CDW.
Filling half of the special sites in the domain wall phase
also explains the densities
$\rho = \frac{1}{2}+\frac{1}{2W}$ of the
$2< \mu/V < 3$ insulating lobe of Fig.~\ref{fig2}(a).

The atomic insulator phases initially persist at small $t/V$, but
the range in chemical potential over which they are stable decreases.
They completely disappear beyond
a critical value of $t/V$.  We note that there is an additional valence-bond
insulator at
$\rho=\frac{1}{4}$ for $W=2$ which
has no atomic counterpart\cite{hou07,bergman13}.
At non-zero $t/V$,
all these insulators are separated by incommensurate superfluid regions.

Quantum phases suggested by these
strong coupling arguments can be precisely determined using the SSE
by measuring the
average density $\rho=\frac{1}{N}\sum_{i}\langle n_{i}\rangle$ and the
superfluid density $\rho_{s}=\langle W^{2}\rangle/4\beta t$,
where $W$ is the winding number counting the net number of times the paths of the particles have wound around the periodic cell\cite{pollock1987,Rousseau2014},
and $\beta$ is the
inverse temperature. Insulating behavior is characterized by
$\rho_s=0$ and
a plateau of $\rho$ representing the persistence of
the atomic limit steps in the chemical potential
to finite $t/V$.
Conversely, the superfluid
has nonzero $\rho_s$ and finite compressibility
$\kappa = \partial \rho/\partial \mu$.  These features are clearly seen
in the SSE results of Fig.~\ref{fig2}(b) for
$W=2$.
A collection of plots like Fig.~\ref{fig2}(b) for different
$t/V$ generates the phase diagram in Fig.~\ref{fig2}(a).

\begin{figure}[htbp]
\centering \includegraphics[width=9cm]{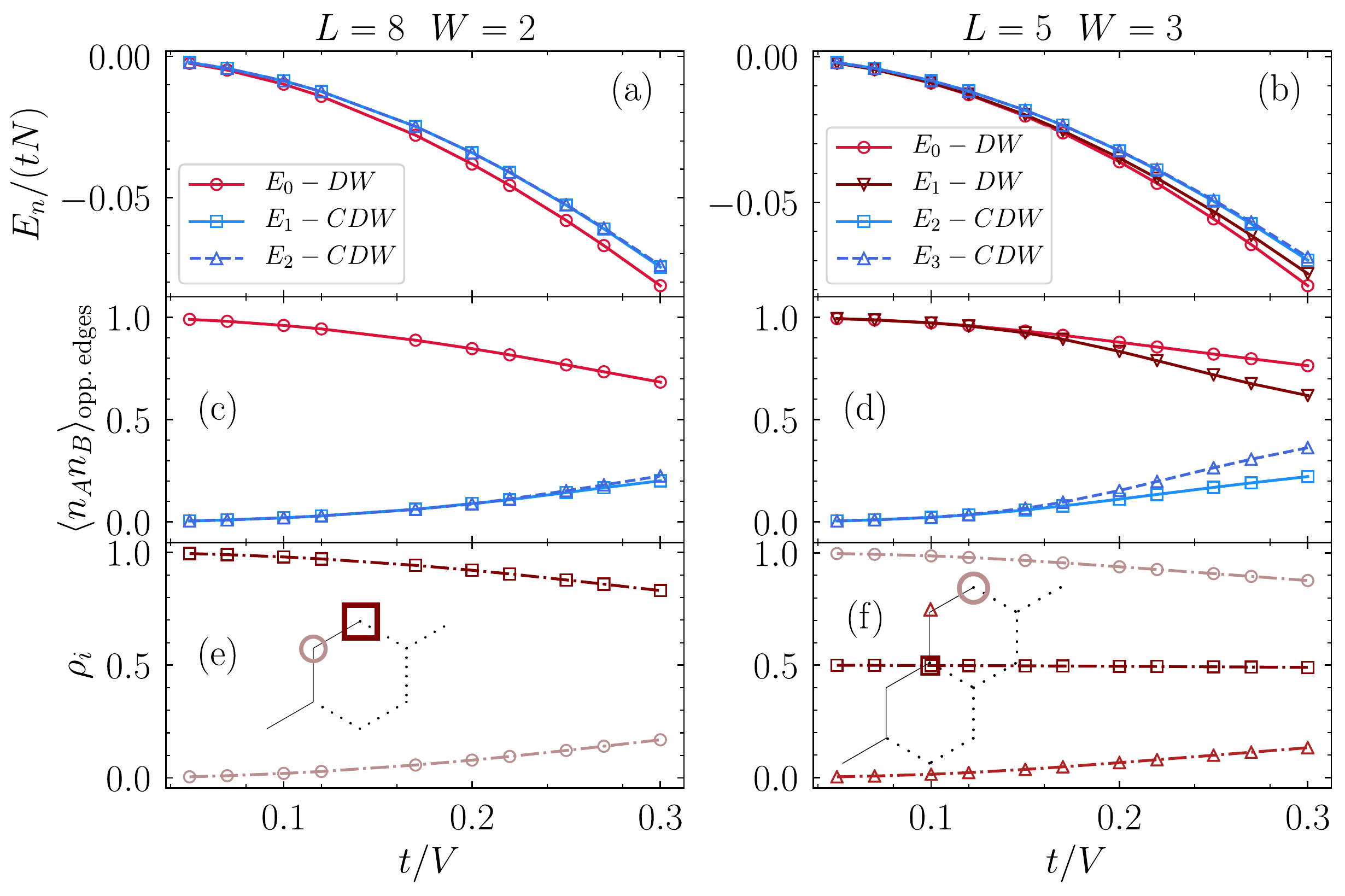} \caption{Several of
the lowest eigenenergies calculated by the ED method: (a), $W=2$; (b), $W=3$. (c) and (d) are the density-density correlations between the opposite-edge sites of the unit cell. Panels (e) and (f) are the corresponding profiles of the
local densities in the ground state. Since the profile is symmetric to
the center of the unit cell, the local densities on only nonequivalent
sites are shown (see the insets).}
\label{fig3}
\end{figure}

We analyze the lifting of the atomic limit $(W+1)$-fold degeneracy
of the $\rho=\frac{1}{2}$ insulator
by the hopping non-perturbatively
on small lattices using ED.
As shown for $W=2$ in Fig.~\ref{fig3}(a), as $t/V$ increases,
$W+1=3$ distinct
eigenenergy curves emerge from the degenerate $t/V=0$ limit.
The site densities and the density-density correlations at Figs.~\ref{fig3}(c) and~\ref{fig3}(e) confirm the lowest
is the unique domain wall phase. When $t$ is non-zero, the single-sublattice CDW states form linear
combinations and in the resulting state superpositions, all sites have average
filling around 0.5. However the domain wall state, because of inversion
symmetry, has two inequivalent sets of sites, half of which have
densities which are low, and half which are high. In the
domain wall phase, both edge sites of the unit cell have high densities,
while one of them is occupied and the other is not in the CDW phase. The
density-density correlations between such sites have distinct values,
which are large for the domain wall phase, and small for the CDW
[see Fig.~\ref{fig3}(e)].
Thus
the presence of two well-separated $\rho_i(t/V)$ trajectories
and large density-density correlation between opposite-edge sites are
`smoking guns'
that the ground state has a domain wall, as already suggested
by the strong coupling argument.

Figures \ref{fig3}(b,d,f) demonstrate the existence of a domain
wall in the $W=3$ ground state.  In Fig.~\ref{fig3}(b), the $W+1=4$-fold
atomic limit degeneracy is lifted by hopping.  In the larger
lattice results, the two lowest states
are nearly degenerate, as the two upper ones~\cite{SM}.  If one averages the $\rho_i$
of the two domain wall states, one finds
two `occupied' sites which have $\rho_i$ large in both states,
two `empty' sites which have $\rho_i$ small in both states, and
two sites which exchange $\rho_i$ small and large.  This leads to
three $\rho_i$ trajectories:  large, small, and approximately
half-filled.
Meanwhile, each of the CDW states has six inequivalent sites,
three high and three low density, which all exchange between
the two degenerate cases.  When averaged, all sites would
therefore have $\rho_i\sim 0.5$.
The density-density correlation between opposite-edge sites in the
domain wall phase is much larger than that in the CDW phase [See
Fig.~\ref{fig3}(d)]. Thus the densities and density-density
correlations observed in Fig.~\ref{fig3}(f) and \ref{fig3}(d) for $W=3$
offer compelling evidences that the ground state manifests a charge
domain wall.

\begin{figure}[htbp]
\centering \includegraphics[width=8.cm]{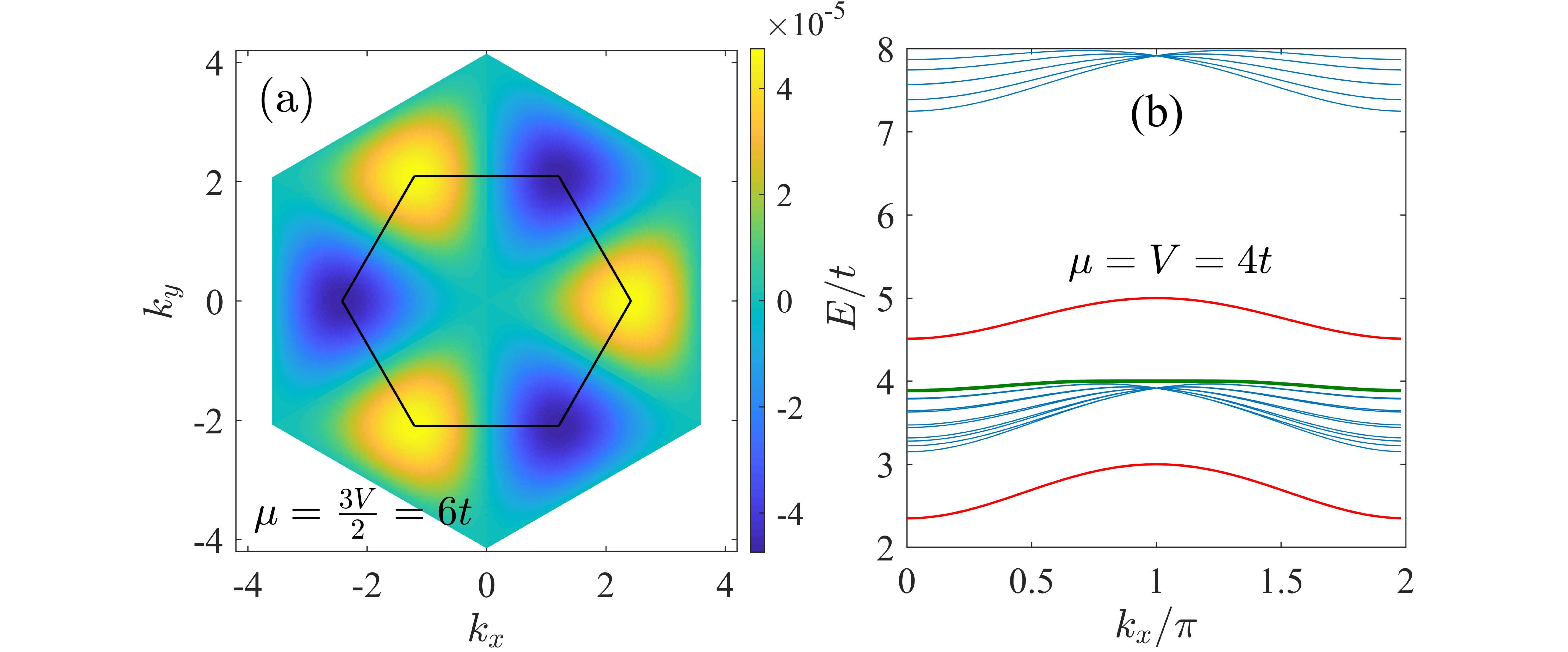} \caption{(a): The Berry curvature associated with the
upper magnon band, which differs from that of the lower band by a sign. The first Brillouin zone is marked by black lines. The
parameters are $\mu=\frac{3V}{2}=6t$. (b) The excitation spectrum on a $W=12$ ribbon with a zigzag domain wall in the middle. The red curves represent states localized near the domain wall. The green curves are two-fold degenerate, and are associated with the zigzag edges. The parameters are $\mu=V=4t$.}
\label{fig4}
\end{figure}

\textit{The Berry curvature and the domain-wall superfluid.-}
Having established the phase diagram and the
existence of domain walls in the $\rho=1/2$ insulator, we now focus on
behavior at the interface. The topological property of bulk CDW phase of hardcore bosons is first investigated.
The Bose-Hubbard Hamiltonian in Eq.(\ref{eq1}) is equivalent to a
spin$-1/2$ $XXZ$ model through a mapping $S^{+}_i=b^{\dag}_i$ and
$S^z_i=n_i-\frac{1}{2}$.   The associated
Holstein-Primakoff transformation yields,
\begin{eqnarray}\label{eq2}
H=&-&t\sum_{\langle i,j\rangle} (a^{\phantom{\dagger}}_{i,A}a^{\phantom{\dagger}}_{j,B}
+a_{i,A}^{\dagger}a_{j,B}^{\dagger}) \\ \nonumber
&+&V\sum_{\langle i,j \rangle} (1-a_{i,A}^{\dagger}a^{\phantom{\dagger}}_{i,A})
a_{i,B}^{\dagger}a^{\phantom{\dagger}}_{i,B} \\ \nonumber
&-&\mu\sum_{i\in A}(1-a_{i,A}^{\dagger}a^{\phantom{\dagger}}_{i,A})-
\mu\sum_{i\in B}a_{i,B}^{\dagger}a^{\phantom{\dagger}}_{i,B}
\end{eqnarray}
where $a^{\dagger}_{i,\alpha},a^{\phantom{\dagger}}_{i,\alpha}$ ($\alpha=A, B$ denoting the
sublattice) are the bosonic creation and annihilation operators\cite{boson}.
In momentum space,
$H=\sum_{\bf k}\psi^{\dagger}_{\bf k}{\cal H}({\bf k})
\psi^{\phantom{\dagger}}_{\bf k}$,
where $\psi_{\bf k}=\{ a^{\phantom{\dagger}}_{A, {\bf k}},
a^{\dagger}_{B, -{\bf k}} \}^{T}$,
and
\begin{eqnarray}\label{eq3}
{\cal H}({\bf k})=\left[
                    \begin{array}{cc}
                      \mu & f({\bf k}) \\
                      f^*({\bf k}) & 3V-\mu \\
                    \end{array}
                  \right],
\end{eqnarray}
with $f({\bf k})=-t(1+e^{-{\rm i}{\bf k}\cdot{\bf a}_1}+e^{-{\rm i}{\bf k}\cdot{\bf a}_2})$.
The magnon band structure, i.e., the excitation of the mapped spin$-1/2$ $XXZ$ model, has two branches: $E^{\pm}_{\bf
k}=\pm(\mu-\frac{3V}{2})+ \epsilon({\bf k})$, where $\epsilon({\bf
k})=\sqrt{(\frac{3V}{2})^2-|f({\bf k})|^2}$ [see Fig.\ref{fig4}(a)]. The
Berry curvature associated with each magnon band is given by
\begin{eqnarray}\label{eq4}
\Omega_{\lambda}({\bf k})=\frac{\partial A_{y}({\bf k})}{\partial k_x}-\frac{\partial A_{x}({\bf k})}{\partial k_y},
\end{eqnarray}
where $A_{i}=-{\rm i}\langle u_{\lambda,{\bf
k}}|\frac{\partial}{\partial k_i}|u_{\lambda,{\bf k}}\rangle$ ($i=x,y$)
is the Berry potential. $\lambda=\pm$ denotes the two magnon
bands\cite{Owerre2016,guo2016,fukui2005}.
The Berry curvature is peaked at the Brillouin zone (BZ)
corners,
Fig.~\ref{fig4}(a),
and is antisymmetric with respect to the inversion center ${\bf
k}=(0,0)$. The sum of the Berry
curvature of each band in the BZ (the Chern number) vanishes
identically. The Berry curvatures for the two $\rho=\frac{1}{2}$ CDW
insulators differ by an overall sign, thus the Berry curvature changes the sign across the
domain wall. Due to bulk-boundary correspondence, there should appear in-gap domain-wall states\cite{semenoff2008,jiang2018}. As shown in Fig.\ref{fig4}(b), there are two such branches associated with the domain wall. One of them is at the bottom of the magnon excitation spectrum, and it corresponds to the superfluid above the $\rho=\frac{1}{2}$ CDW insultor, which is localized near the domain wall.

\begin{figure}[htbp]
\centering \includegraphics[width=7.cm]{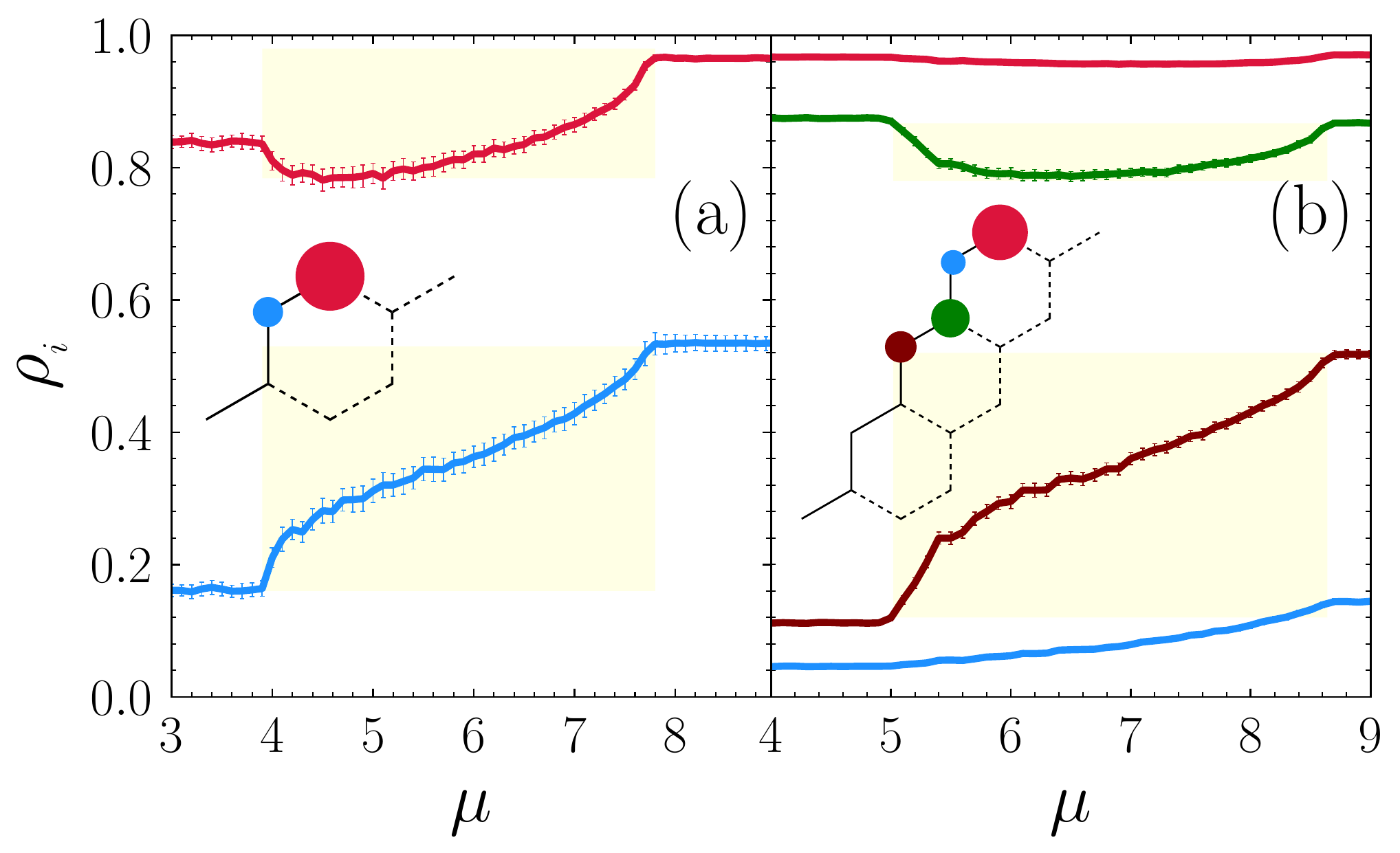} \caption{The local
densities as a function of the chemical potential: (a), $W=2$ and
$t/V=0.3$; (b), $W=4$ and $t/V=0.25$. Due to the geometric symmetries, only
nonequivalent sites are shown. Here the length of the ribbons is
$L=24$.}
\label{fig5}
\end{figure}

For the density $\rho=\frac{1}{2}$, the domain wall is thus formed by a quasi-1D region with negligible occupancy, confined by robust CDW phases. As the chemical potential is further increased, bosons are preferentially added to the empty sites forming the domain wall, since they only interact with two occupied neighbors in contrast with three neighbors of an empty CDW site. A single extra boson can hop freely along the chain, lowering
its kinetic energy without changing the interacting energy.  Additional
added particles behave effectively as interacting bosons in 1D, which
condense to superfluid transport down the domain wall. The values of the superfluid density,
Fig.~\ref{fig2}(b), follow a dome shape, and are maximal at
density $\sim \frac{1}{2}+\frac{1}{4W}$ when half of such
empty sites are occupied. After the domain wall is full, the superfluid
vanishes, and the system becomes a $\rho=\frac{1}{2}+\frac{1}{2W}$
insulator. Figure \ref{fig5} shows the
density for $W=2, 4$.  The presence of two sets of well-separated
local traces is consistent with
the $\rho=\frac{1}{2}$ insulator being a domain-wall
phase.  (See discussion of Fig.~\ref{fig3}(e).)

While
the local density of the sites on the domain wall with small occupation
increases, that of the high-occupancy sites
first decreases even as $\mu$ grows.
This anomalous behavior is a signature of
the flow of bosons onto the domain wall and
the appearance of a superfluid
localized near the domain wall.
It is noteworthy that the domain wall superfluid
coexists with diagonal (density) order:
the zigzag honeycomb nanoribbon
realizes an
exotic one-dimensional supersolid\cite{batrouni2006}. The emergence of
this superfluid is a manifestation of change in topological properties
when crossing the domain wall.

\begin{figure}[htbp]
\centering \includegraphics[width=7.cm]{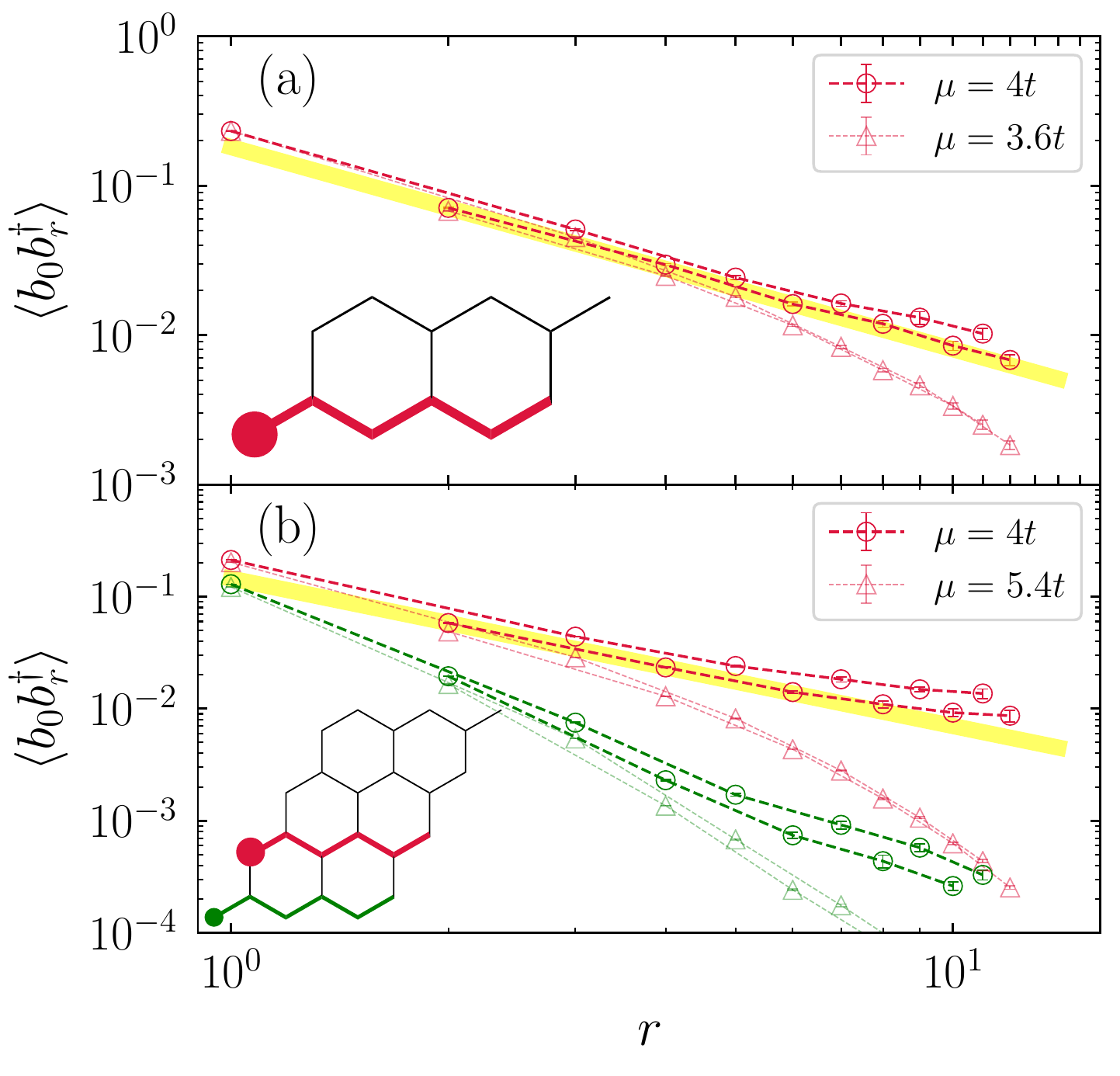}
\caption{The single-particle correlator
   $\langle b^{\dagger}_{0} b_{r}\rangle$
(a), $W=2$; (b), $W=4$. The circle symbols on the
inset geometries mark the reference site $r=0$. The correlators
are calculated along the thick zigzag lines, and only nonequivalent
lines are shown.
The thick yellow lines are plotted as guides to algebraic behavior. Up triangles connected by dotted lines refer to the insulating regime with $\rho = \frac{1}{2}$. Parameters are the same as those in Fig.~\ref{fig5}.
}
\label{fig6}
\end{figure}

To verify the localization of the superfluid near the domain
wall, we show the single-particle correlator $\langle
b^{\dagger}_{0}b^{\phantom{\dagger}}_{r}\rangle$
in Fig.~\ref{fig6}. The correlator along the
zigzag chain on the domain wall
is slower than a power-law
decay with distance, which is characteristic of a gapless quasi-1D superfluid. In
contrast, the excitation is gapped for the $\rho=\frac{1}{2}$
domain-wall insulator, and the correlator decays exponentially. As
one moves away from the domain wall, the correlator becomes increasingly
short-ranged, and $\rho_s$ decreases.  For wide ribbons, the superfluid
density decays exponentially with the distance away from the domain
wall\cite{SM}.

\textit{Conclusions.-}
We studied hardcore bosons on zigzag edge honeycomb lattice ribbons
using exact simulations. The phase diagram
contains superfluid and insulating phases and, remarkably,
at $\rho=\frac{1}{2}$ filling the ground state contains
a charge domain wall rather than occupation of a single sublattice.
This `self-organized domain
wall' separates CDW regions with opposite Berry curvature,
and supports superfluid transport in coexistence with
diagonal (density) order.
Our results demonstrate that honeycomb ribbons
provide a concrete geometry for the observation of bosonic topological
phenomenon.

This physics can be explored experimentally.
Cold atoms in optical lattices provide a well-established means
to emulate the Bose-Hubbard model\cite{jaksch98}, large values of
$U$ which achieve the hard-core limit can be attained,
and the honeycomb geometry has been
generated~\cite{Polini2013, zhu2007, soltanpanahi2011}. The use of a synthetic dimension may also provide a simple realization of honeycomb ribbons\cite{suszalski2016,ozawa2019}. New
tools based on quantum gas microscopes allow observation of
the density profile at the level of individual
atoms\cite{Gross2017,bloch2012,bakr2009,sherson2010}, and hence
direct comparison with our real-space measurements.  The Berry
curvature can also be obtained via interferometric
techniques\cite{Duca2015,Flaschner2016}.

\textit{Acknowledgments.-}
The authors thank G. Batrouni, G. Chen and F. Nori for helpful discussions. H.G. acknowledges support from the NSFC grant No.~11774019.
X.Z. and S.F. are supported by the National Key Research and Development Program of China under Grant No. 2016YFA0300304, and NSFC under Grant Nos. 11574032 and 11734002. R.M.~acknowledges support from NSAF-U1530401, and from NSFC Grant No. 11674021  and No. 11851110757. R.T.S.~was supported by DOE grant DE-SC0014671.

\nocite{*}
\bibliography{zigzag_ref}

\begin{thebibliography}{48}
\bibitem [{bos()}]{boson}%
  \BibitemOpen
  \href@noop {} {}\bibinfo {note} {Here the bosons are used to express the
  spin$-1/2$ operators, and they are distinct from hardcore bosons in
  Eq.(1).}\BibitemShut {Stop}%
\end{thebibliography}
\appendix
\renewcommand{\thefigure}{S\arabic{figure}}

\section{1. Band structure}

\begin{figure}[htbp]
\centering \includegraphics[width=9.cm]{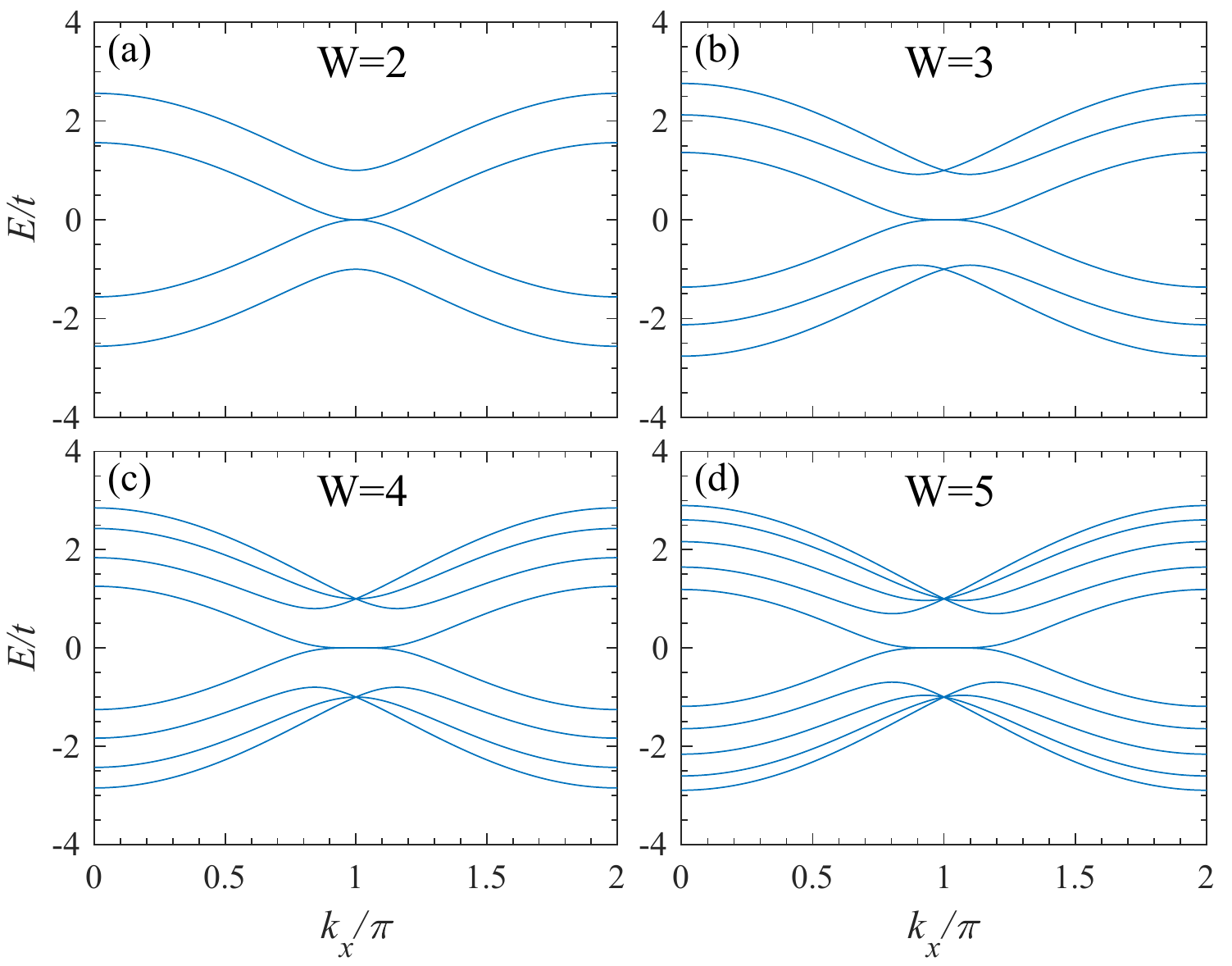} \caption{ The band
structures of the honeycomb lattice ribbons with the widths: (a), $W=2$;
(b), $W=3$; (c), $W=4$; (d), $W=5$. }
\label{afig1}
\end{figure}

The band structure of
honeycomb lattice ribbons can be determined analytically.
One fermion has exactly
the same energy as one hardcore boson due to the absence of exchange
statistics. For $W=2$, the Hamiltonian in the momentum space is,
\begin{eqnarray}\label{eqa1}
 H_{2}(k_x)=
\left[
  \begin{array}{cccc}
    0 & -t\gamma_{k} & 0 & 0 \\
    -t\gamma^*_{k} & 0 & -t & 0 \\
    0 & -t & 0 & -t\gamma_{k} \\
    0 & 0 & -t\gamma^*_{k} & 0 \\
  \end{array}
\right],
\end{eqnarray}
with $\gamma_{k}=1+e^{ik_x}$.
The energy spectrum contains four branches, $E_{i}=(\pm 1\pm
\sqrt{9+8\cos k_x})t/2$ ($i=1,2,3,4.$). The band bottom is located at
$k_x=0$, and the corresponding eigenvalue is $-(1+\sqrt{17})t/2$ (See Fig.~\ref{afig1}). Thus
the lower boundary of the phase diagram, where the density
first begins to become nonzero, is a straight line
$\mu/V=-\frac{1+\sqrt{17}}{2} t/V$.

Under a particle-hole transformation $b^{\dagger}_{i}\rightarrow h_{i}$,
the Hamiltonian, Eq.~(1) of the main text, becomes,
\begin{eqnarray}\label{eqa2}
H=&-&t\sum_{\langle i,j\rangle} (h_i^{\dagger}
h^{\phantom{\dagger}}_{j}+\text{H.c.})
+V\sum_{\langle i,j \rangle} n^{h}_i n^{h}_j \\ \nonumber
  &-&(3V-\mu) \sum_{i\in bulk} n^{h}_i-(2V-\mu) \sum_{i\in edge} n^{h}_i+E_{0},
\end{eqnarray}
where $n^{h}_i=h_i^{\dagger}h_i$ is the hole number operator, and
$E_{0}=\frac{3}{2}VN-LV-\mu N$ ($N=2L \times W$ the total number of
sites) is a constant.  In momentum space, the hole Hamiltonian is
\begin{eqnarray}\label{eqa3}
 H_{2}^{h}(k_x)=
\left[
  \begin{array}{cccc}
    -2V & -t\gamma_{k} & 0 & 0 \\
    -t\gamma^*_{k} & -3V & -t & 0 \\
    0 & -t & -3V & -t\gamma_{k} \\
    0 & 0 & -t\gamma^*_{k} & -2V \\
  \end{array}
\right].
\end{eqnarray}
The energy spectrum $E_{1,2}=-\frac{t}{2}-\frac{5V}{2}\pm
\frac{1}{2}P_{k,+}; E_{3,4}=\frac{t}{2}-\frac{5V}{2}\pm
\frac{1}{2}P_{k,-}$ with $P_{k,\pm}=\sqrt{9t^2+8t^2\cos k_x+V^2\pm
2tV}$. The band bottom determines the upper boundary of the phase
diagram, which is described by the curve
$\frac{\mu}{V}=\frac{1}{2}\frac{t}{V}+\frac{1}{2}\sqrt{17\frac{t^2}{V^2}+2\frac{t}{V}+1}+\frac{5}{2}$.

For the $W=3$ case, the Hamiltonian is,
\begin{eqnarray}\label{eqa4}
H_{3}(k_x)=
\left[
  \begin{array}{ccc}
    h_{1} (k_x)& h_{2} & 0 \\
    h^{\dagger}_{2} & h_{1} (k_x) & h_{2} \\
    0 & h^{\dagger}_{2} & h_{1} (k_x) \\
  \end{array}
\right],
\end{eqnarray}
where
\begin{eqnarray}\label{eqa6}
h_{1}(k_x)=\left(
                  \begin{array}{cc}
                    0 & -t\gamma_{k} \\
                    -t\gamma^{*}_{k} & 0 \\
                  \end{array}
                \right), h_{2}=\left(
                                 \begin{array}{cc}
                                   0 & 0 \\
                                   -t & 0 \\
                                 \end{array}
                               \right). \nonumber
\end{eqnarray}
The lower boundary is described by $\mu/V=-2.76 \,t/V$.

For $W=4$, the Hamiltonian is
\begin{eqnarray}\label{eqa5}
H_{4}(k_x)=
\left[
  \begin{array}{cccc}
    h_{1} (k_x)& h_{2} & 0 & 0 \\
    h^{\dagger}_{2} & h_{1} (k_x) & h_{2} & 0 \\
    0 & h^{\dagger}_{2} & h_{1} (k_x) & h_{2} \\
    0 & 0 & h^{\dagger}_{2} & h_{1} (k_x) \\
  \end{array}
\right].
\end{eqnarray}
From the eigenvalue of the band bottom, we have $\mu/V=-2.851 \,t/V$
describing the phase boundary for $\rho=0$. The phase boundary for
$\rho=1$ can be obtained numerically in the hole representation.

\section{2. The phase diagrams for several other widths}

\begin{figure}[htbp]
\centering \includegraphics[width=9.cm]{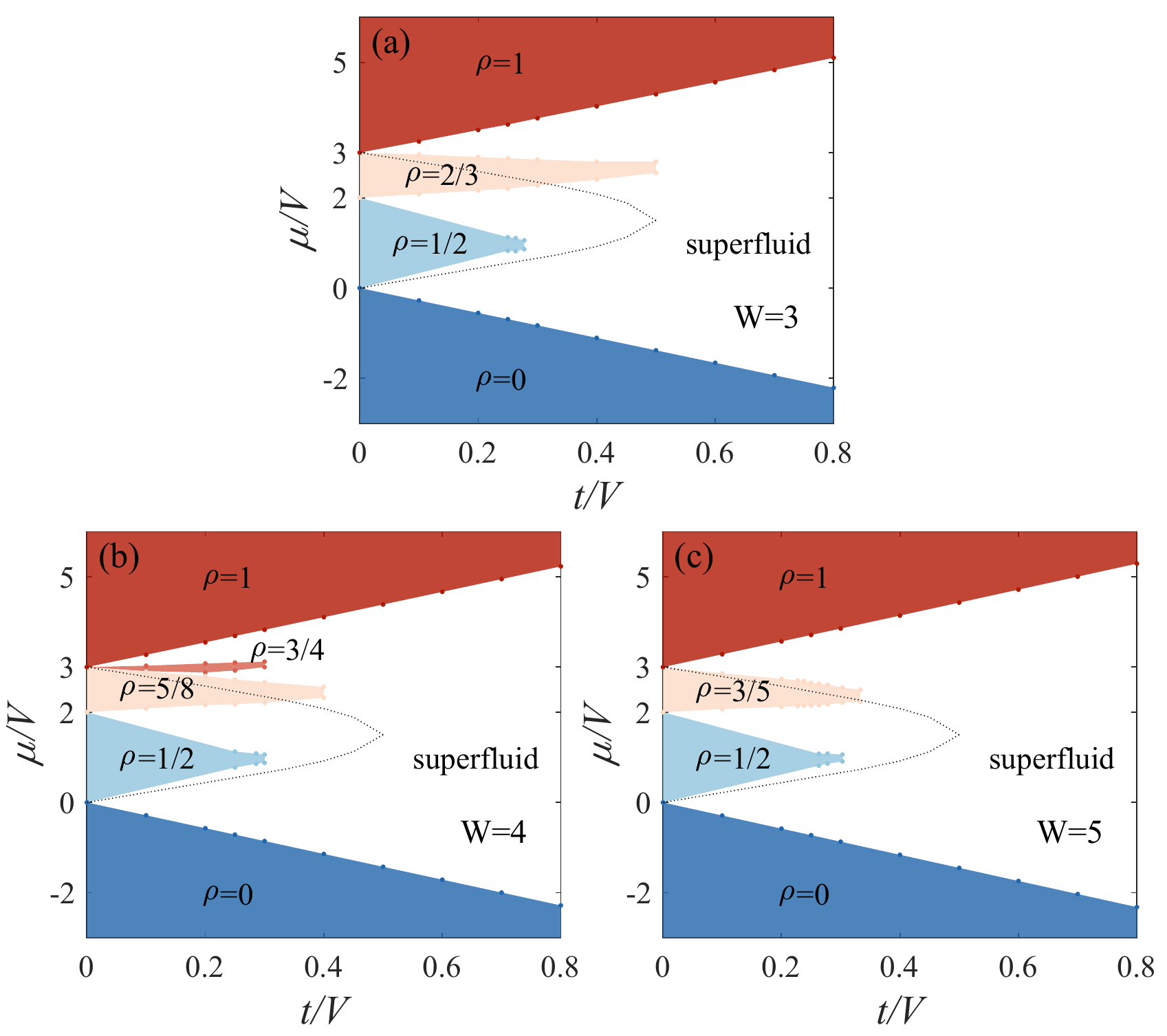} \caption{
Phase diagrams of $W=3,4,5$ honeycomb lattice ribbons.
Dotted lines enclose the $\rho=\frac{1}{2}$ CDW
insulator of hardcore bosons on a periodic two-dimensional honeycomb
lattice. Ribbon
length $L=24$.}
\label{afig2}
\end{figure}

The phase diagrams of the ribbons with several other widths are shown in Fig.\ref{afig2}, which contain superfluid regions and insulator phases at specific fillings. Compared to that of the periodic honeycomb lattice, the regions of the $\rho=\frac{1}{2}$ insulators are greatly reduced, and there appear new insulating regions at the filling $\rho=\frac{1}{2}+\frac{1}{2W}$.  There is an additional valence-bond
insulators at $\rho=\frac{3}{4}$ for $W=4$ which
has no atomic counterpart.

\section{3. Strong Coupling}
\begin{figure}[htbp]
\centering \includegraphics[width=8.cm]{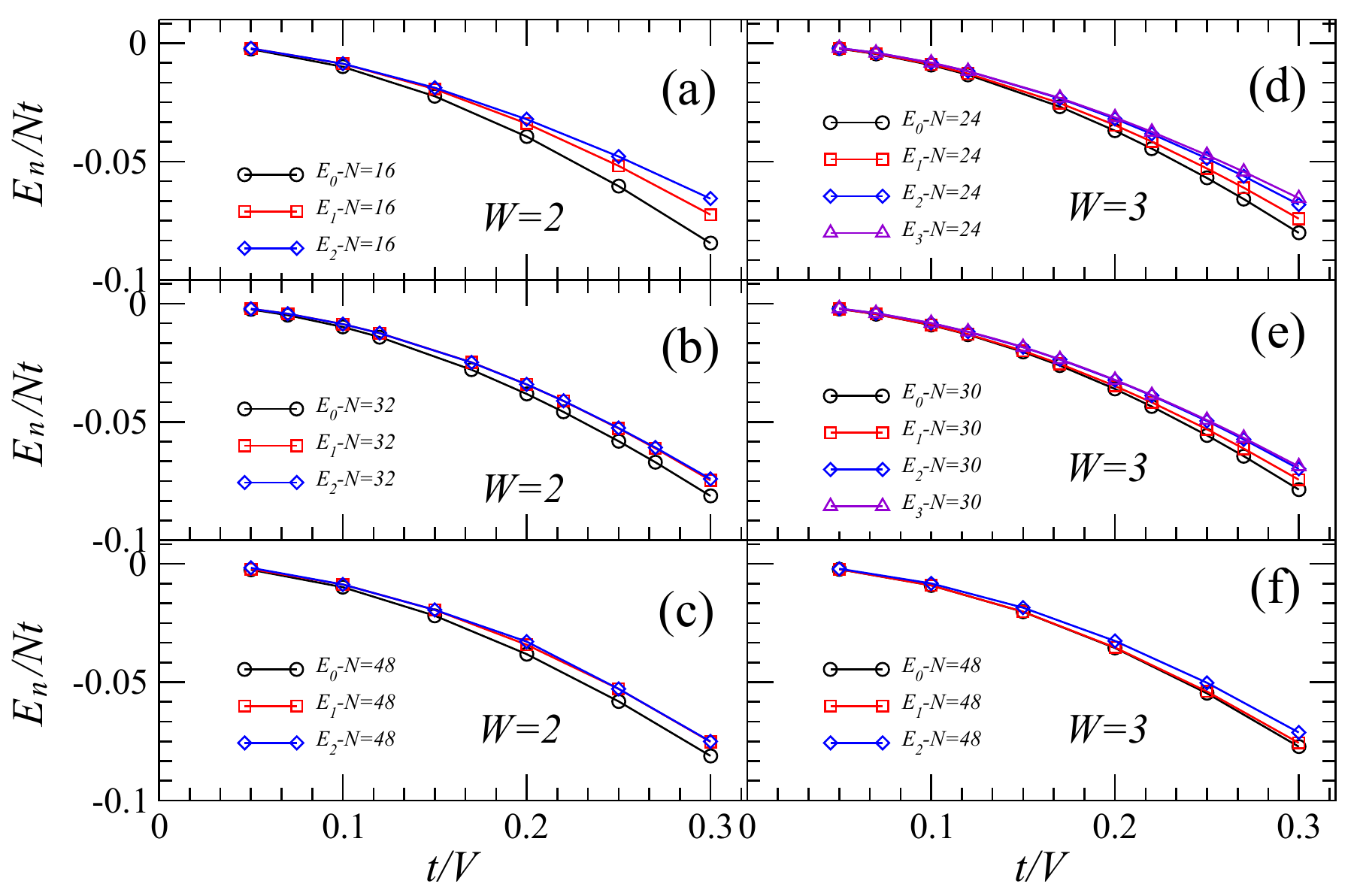} \caption{Several of
the lowest eigenenergies calculated by the ED method: (a), $W=2$, $L=4$; (b), $W=2$, $L=8$; (d), $W=3$, $L=4$; (e), $W=3$, $L=5$. The three lowest eigenenergies from the DMRG methods on $N=48$ sites: (c), $W=2$; (f), $W=3$.}
\label{afig3}
\end{figure}

\begin{figure}[htbp]
\centering \includegraphics[width=7.cm]{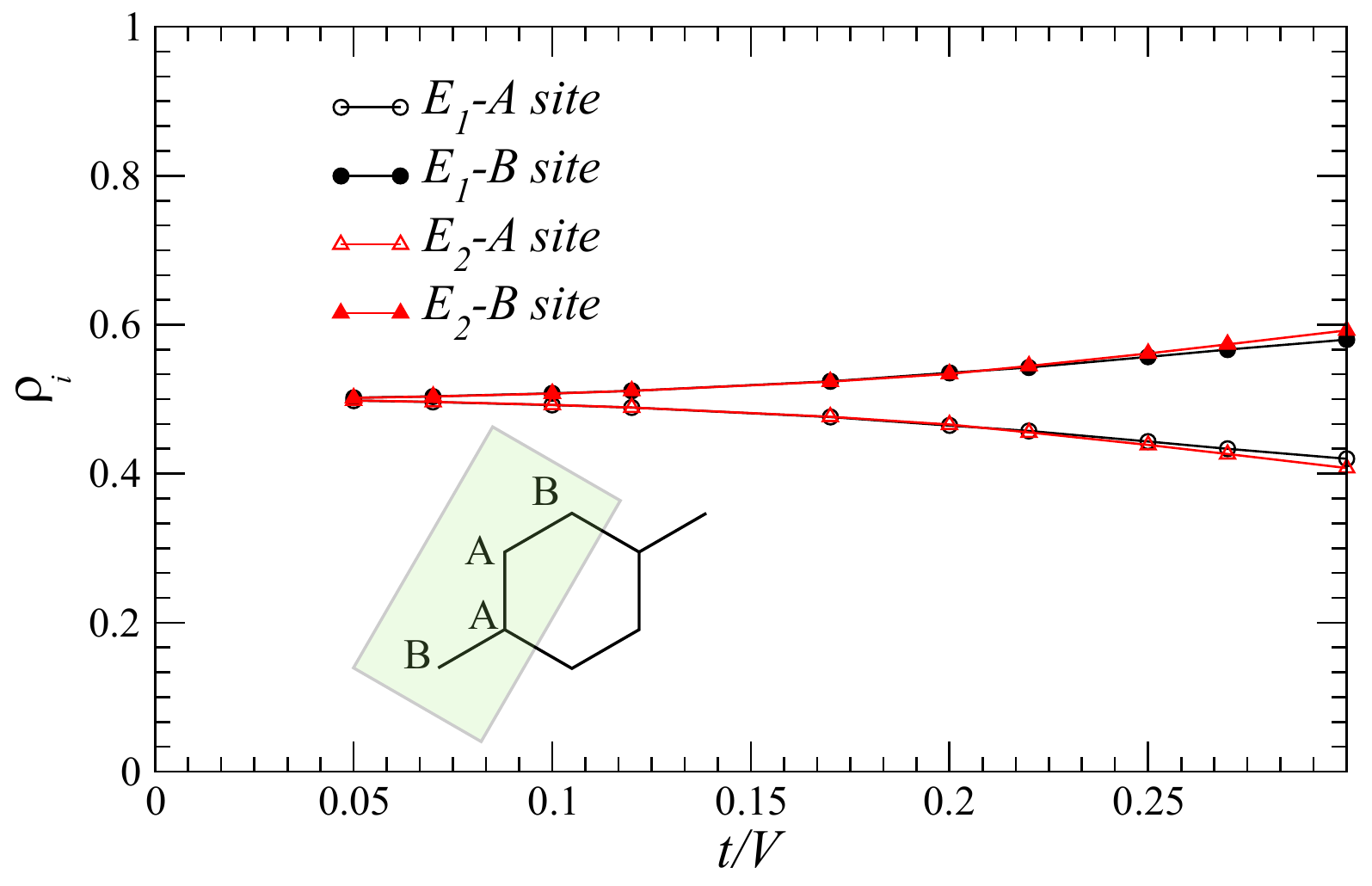} \caption{The profiles of the
local densities in the first- and second-excited states, which are linear combinations of the two single-sublattice CDW states. The lattice size is $W=2$ and $L=8$. Inset shows the unit cell of the $W=2$ ribbon and equivalent sites, due to inversion point group symmetry.}
\label{afig4}
\end{figure}

The Exact Diagonalization and Density Matrix Renormalization methods
can be used to obtain further details of the evolution of the low energy
spectrum and site densities away from the atomi ($t=0$) limit.
The $W+1$-fold degeneracy of the atomic limit $\rho=\frac{1}{2}$ insulator is lifted by the hopping. For the $W=2$ case, the ground state is the unique domain wall phase. The first- and second- excited states are linear combinations of the single-sublattice CDW states to preserve the symmetries of the Hamiltonian. As shown in Fig.~\ref{afig3}, the $N=16$ site ED
eigenenergies of the
two mixed single-sublattice CDW states lie above the domain wall
state and are slightly split by finite size
effects.  This finite size splitting is significantly
reduced in the $N=32$ site ED and the $N=48$ site DMRG calculations. We plot the site densities of the
mixed CDW states in Fig.~\ref{afig4}, which are the first- and second-excited states. All sites have average filling around $0.5$. As the value of $t/V$ increases, the density difference between the two inequivalent sites becomes larger due to stronger quantum fluctuations.

\begin{figure}[htbp]
\centering \includegraphics[width=7.cm]{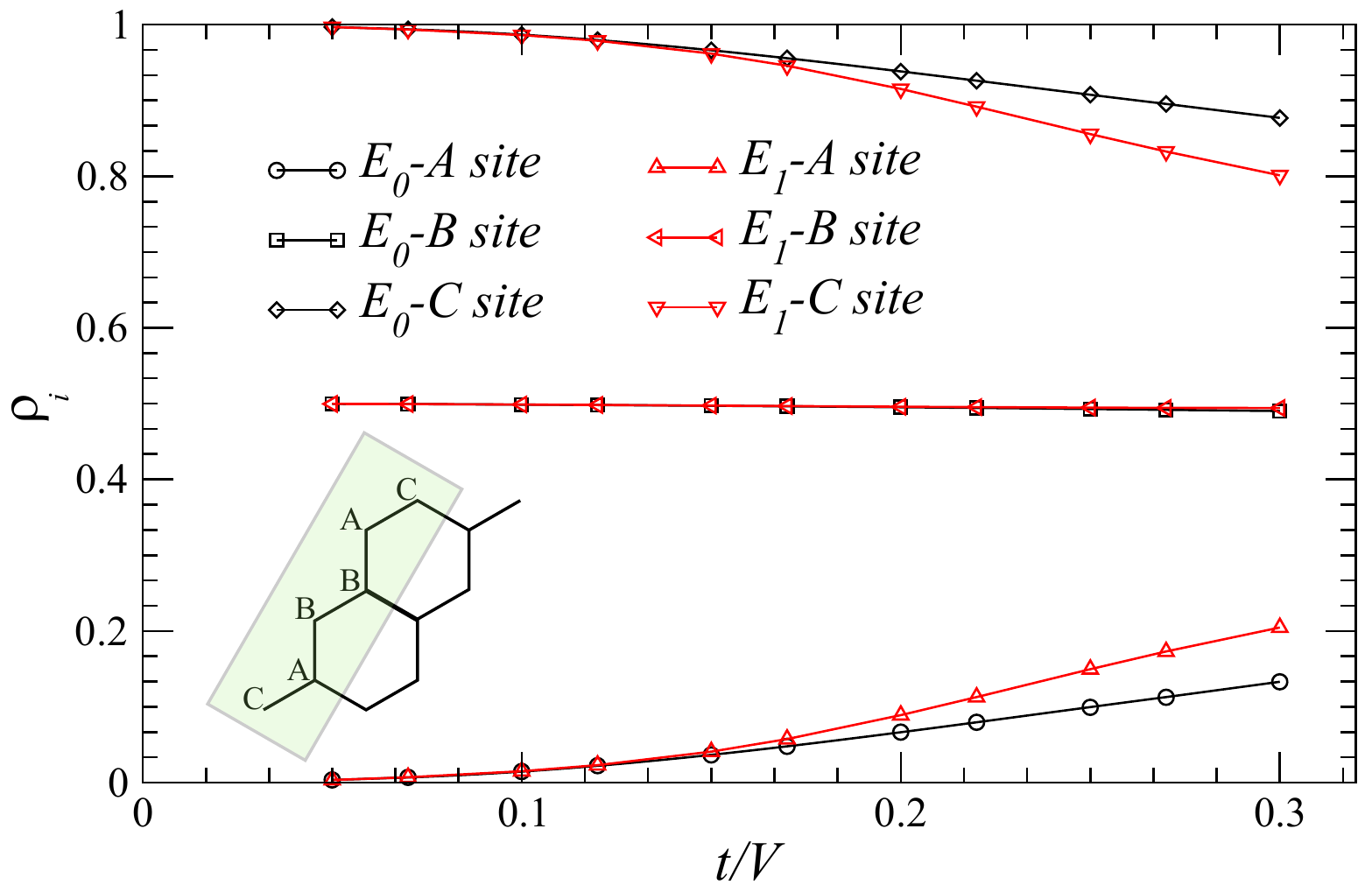} \caption{The profiles of the
local densities in the ground- and first-excited states, which are linear combinations of the two domain-wall phases. The lattice size is $W=3$ and $L=5$.}
\label{afig5}
\end{figure}

\begin{figure}[htbp]
\centering \includegraphics[width=7.cm]{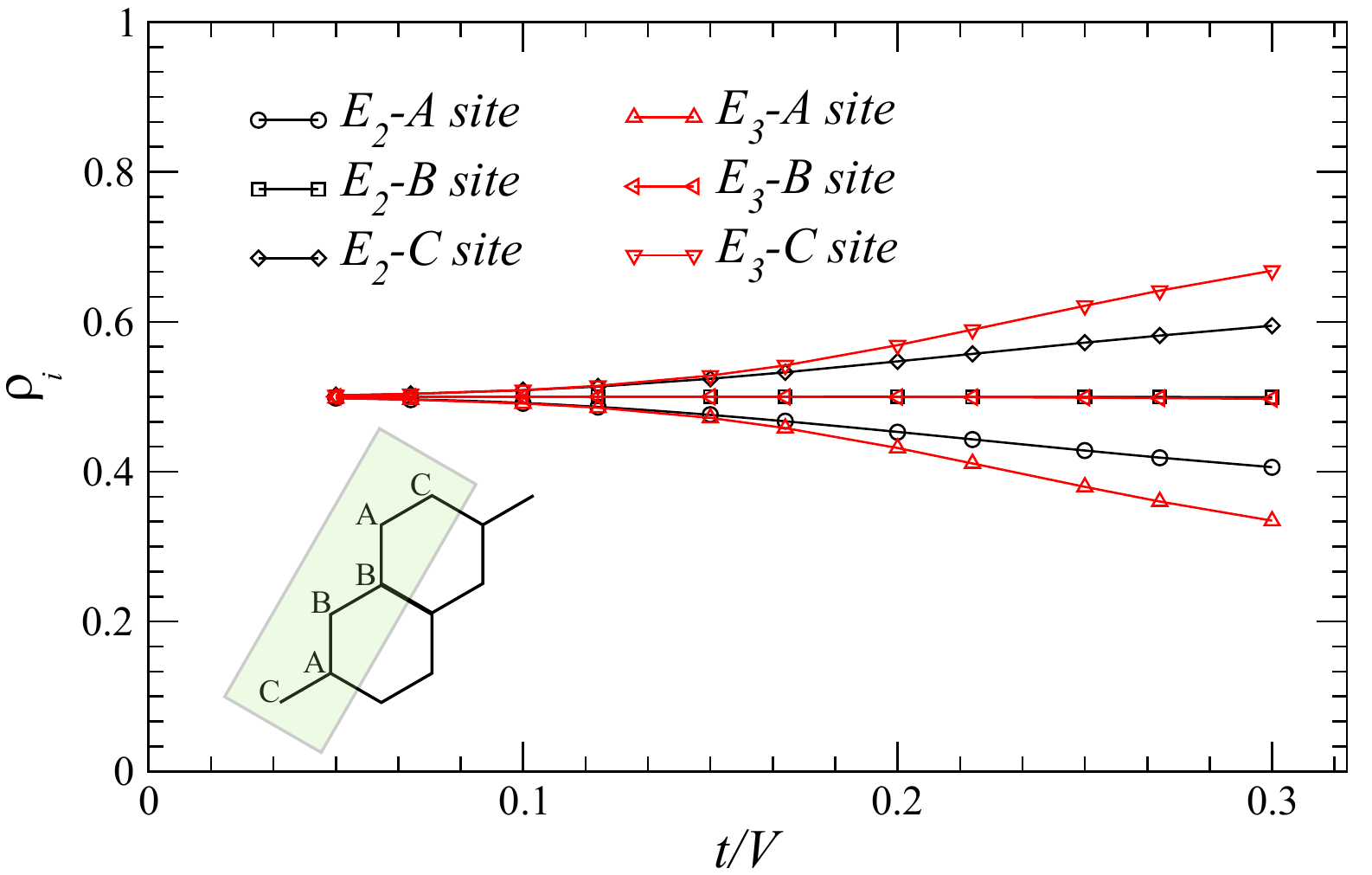} \caption{The profiles of the
local densities in the second- and third-excited states, which are linear combinations of the two single-sublattice CDW states. The lattice size is $W=3$ and $L=5$.}
\label{afig6}
\end{figure}

For the $W=3$ case, the lowest four eigenenergies are grouped into two sets, each of which contains two mixed domain-wall or CDW states. Due to the finite-size effect, the eigenenergies are split. As the lattice size is increased (see Fig.~\ref{afig3}), the splitting of the lower (upper) two eigenenergies is significantly reduced, suggesting they are degenerate in the thermodynamic limit. The site densities of the lower (upper) two states are plotted in Fig.~\ref{afig5}
(Fig.~\ref{afig6}), which are consistent with those of the domain-wall (CDW) phases. Thus the domain-wall phase is the ground state of the $W=3$ ribbon.

\section{4. The Berry curvature}
The model in Eq.~(1) is equivalent to a spin$-1/2$ $XXZ$ model through a mapping $S^{+}_i=b^{\dag}_i$ and $S^z_i=n_i-\frac{1}{2}$,
\begin{eqnarray}\label{eqb1}
H=&-&t\sum_{\langle i,j\rangle} (S_i^{+}S_j^{-}+S_i^{-}S_j^{+}) \\ \nonumber
&+&V\sum_{\langle i,j \rangle} (S^z_i+\frac{1}{2})(S^z_j+\frac{1}{2})
-\mu\sum_{i}(S^z_i+\frac{1}{2}).
\end{eqnarray}

\begin{figure}[htbp]
\centering \includegraphics[width=4.5cm]{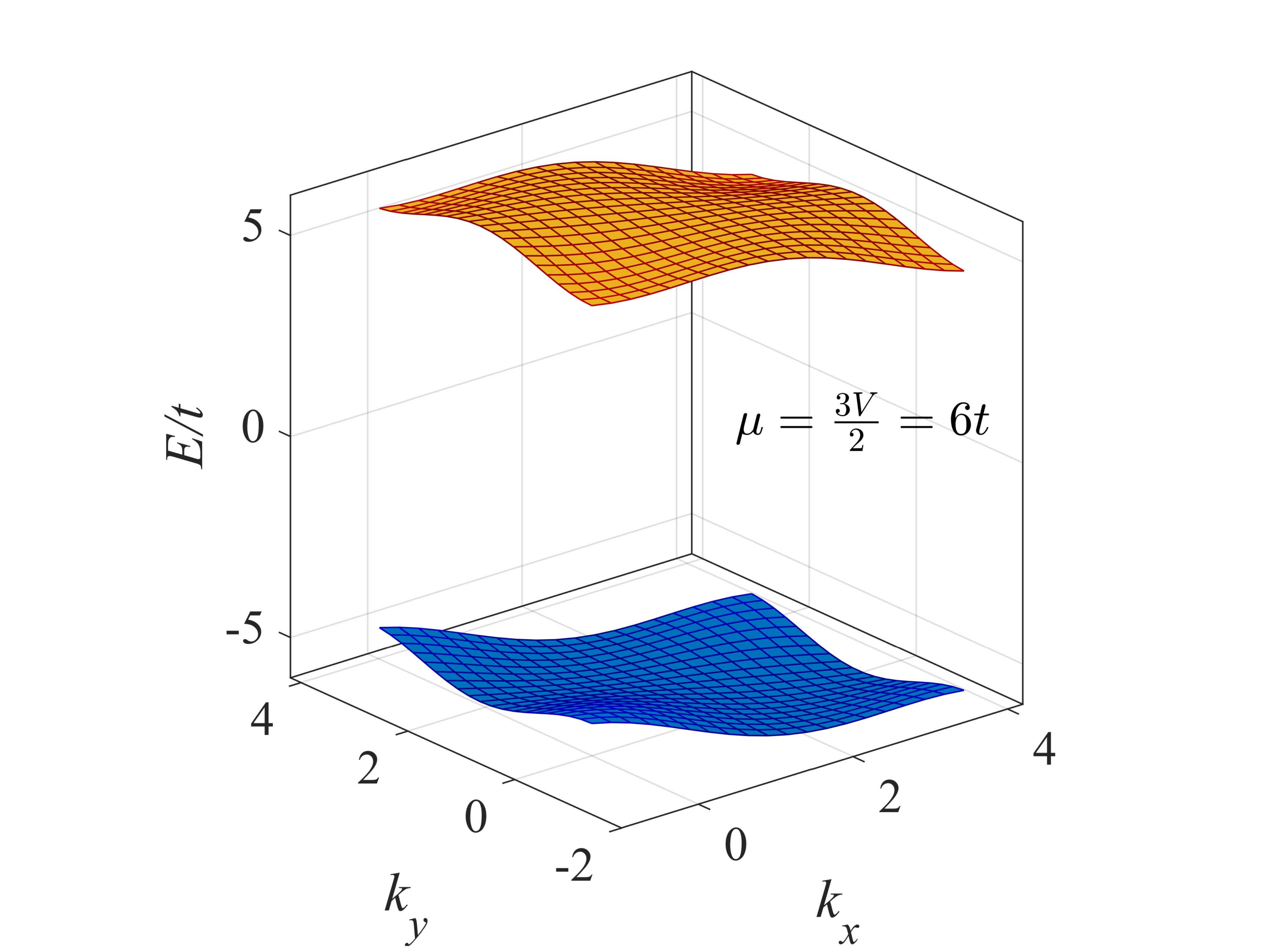} \caption{(a) The magnon band structure. The
parameters are $\mu=\frac{3V}{2}=6t$, when $E^{+}, E^{-}$ are identical and we plot $-E^{-}$ to display it.}
\label{afig7}
\end{figure}

Using the Holstein-Primakoff transformation, the spin operators are
expressed in terms of bosonic creation and annihilation operators. The
honeycomb lattice is bipartite. The transformation on sublattice $A$
is defined as
\begin{align}\label{eqb2}
S^+_{A,i}=(\sqrt{2S-a^{\dagger}_{i,A}a^{\phantom{\dagger}}_{i,A}})a^{\phantom{\dagger}}_{i,A}, \\ \nonumber
S^-_{A,i}=a^{\dagger}_{i,A}(\sqrt{2S-a^{\dagger}_{i,A}a^{\phantom{\dagger}}_{i,A}})  \\ \nonumber
S^z_{A,i}=S-a^{\dagger}_{i,A}a^{\phantom{\dagger}}_{i,A}.
\end{align}
Conversely, on sublattice $B$, the spin is in the opposite direction for
antiferromagnetic order. Thus the spin operators are defined as
\begin{align}\label{eqb3}
S^+_{B,i}=a^{\dagger}_{i,B}(\sqrt{2S-a^{\dagger}_{i,B} a^{\phantom{\dagger}}_{i,B}}), \\ \nonumber
S^-_{B,i}=(\sqrt{2S-a^{\dagger}_{i,B} a^{\phantom{\dagger}}_{i,B}}) a^{\phantom{\dagger}}_{i,B}  \\ \nonumber
S^z_{B,i}=a^{\dagger}_{i,B} a^{\phantom{\dagger}}_{i,B}-S.
\end{align}
Expanding the square root in Eq.(\ref{eqb3}) in powers of $1/S$, the
zeroth order terms are kept in the linear spin-wave theory.
Then the bosonic tight binding Hamiltonian becomes
\begin{eqnarray}\label{eqb4}
H=&-&t\sum_{\langle i,j\rangle} (a^{\phantom{\dagger}}_{i,A}a^{\phantom{\dagger}}_{j,B}+a_{i,A}^{\dagger}a_{j,B}^{\dagger}) \\ \nonumber
&+&V\sum_{\langle i,j \rangle} (1-a_{i,A}^{\dagger}a^{\phantom{\dagger}}_{i,A})a_{j,B}^{\dagger}a^{\phantom{\dagger}}_{j,B} \\ \nonumber
&-&\mu\sum_{i\in A}(1-a_{i,A}^{\dagger}a^{\phantom{\dagger}}_{i,A})-\mu\sum_{i\in B}a_{i,B}^{\dagger}a^{\phantom{\dagger}}_{i,B}.
\end{eqnarray}
We ignore the four-operator terms and a constant. Under a Fourier
transformation, $H=\sum_{\bf
k}\psi^{\dagger}_{\bf k}{\cal H}({\bf k})\psi^{\phantom{\dagger}}_{\bf k}$, where $\psi_{\bf
k}=\{ a^{\phantom{\dagger}}_{A, {\bf k}}, a^{\dagger}_{B, -{\bf k}} \}^{T}$ is the basis,
and
\begin{eqnarray}\label{eqb5}
{\cal H}({\bf k})=\left[
                    \begin{array}{cc}
                      \mu & f({\bf k}) \\
                      f^*({\bf k}) & 3V-\mu \\
                    \end{array}
                  \right]
\end{eqnarray}
with $f({\bf k})=-t(1+e^{-{\rm i}{\bf k}\cdot{\bf a}_1}+e^{-{\rm i}{\bf k}\cdot{\bf
a}_2})$ [${\bf a}_1=(\sqrt{3},0),{\bf a}_2=(\sqrt{3}/2,3/2)$ are the
primitive vectors]. To diagonalize the above Hamiltonian, we consider
the following non-Hermitian matrix,
\begin{eqnarray}\label{eqb6}
\left[
                    \begin{array}{cc}
                      \mu & f({\bf k}) \\
                      -f^*({\bf k}) & -3V+\mu \\
                    \end{array}
                  \right].
\end{eqnarray}
The eigenvalues are given by $E^{\pm}_{\bf k}=\mu-\frac{3V}{2}\pm
\epsilon({\bf k})$ with $\epsilon({\bf
k})=\sqrt{(\frac{3V}{2})^2-|f({\bf k})|^2}$. The eigenvector matrix is
\begin{eqnarray}\label{eqb7}
{\cal U}_{\bf k}=\left[
                    \begin{array}{cc}
                     \cosh \theta_{\bf k}e^{i\phi_{\bf k}} & -\sinh \theta_{\bf k} \\
                      -\sinh \theta_{\bf k} & \cosh \theta_{\bf k}e^{-i\phi_{\bf k}} \\
                    \end{array}
                  \right],
\end{eqnarray}
where $\sinh 2\theta_{\bf k}=\frac{|f({\bf k})|}{\epsilon({\bf k})}$, $\tan \phi_{\bf k}=\frac{Im f({\bf k})}{Re f({\bf k})}$. The first (second) column is the eigenvector $u_{+,{\bf k}}$ ($u_{-,{\bf k}}$) corresponding to $E^{+}_{\bf k}$ ($E^{-}_{\bf k}$). The Hamiltonian is thus diagonalized by the transformation: ${\cal U}^{\dagger}_{\bf k}{\cal H}({\bf k}){\cal U}_{\bf k}=\textrm{diag}(E^{+}_{\bf k}, -E^{-}_{\bf k})$. In Fig.\ref{afig7}, we plot the magnon band structure at $\mu=\frac{3V}{2}=6t$, which consist of two gapped branches.

\begin{figure}[htbp]
\centering \includegraphics[width=5.5cm]{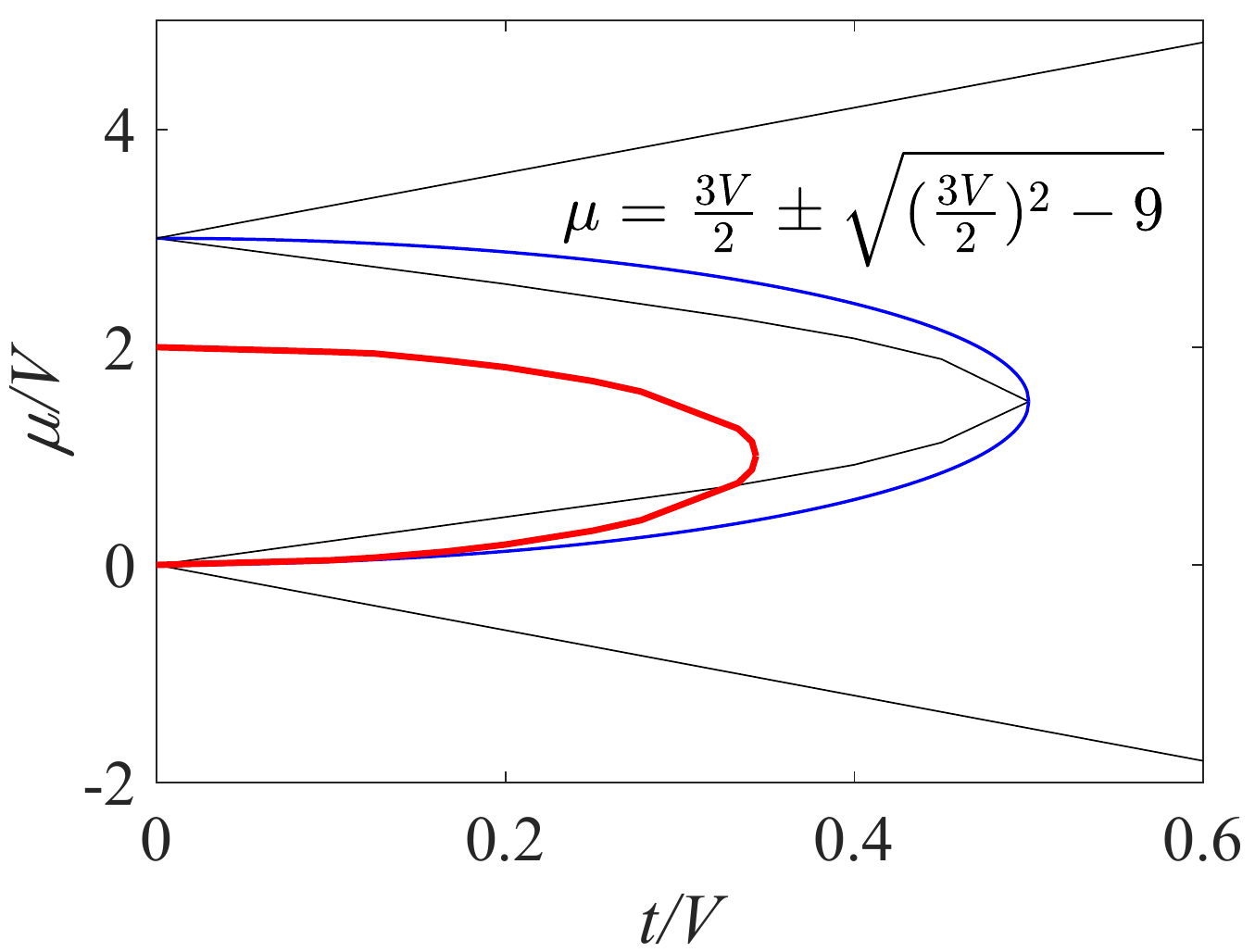} \caption{(a) The phase diagram of the Bose-Hubbard model on periodic honeycomb lattice and zigzag ribbon. The black (blue) curves are the phase boundaries from the QMC method (the spin-wave approximation) on periodic honeycomb lattice. The red curves are from the spin-wave approximation for a zigzag ribbon with the width $W=12$.}
\label{afig8}
\end{figure}

The above magnon bands of the mapped spin$-1/2$ XXZ model correspond to the excitation spectrum above the CDW insulator of the Bose-Hubbard model, which is the superfluid. When the spectrum becomes gapless, i.e., $E^{\pm}_{\bf k}=0$, superfluid begins to appear in the system. Thus $E^{\pm}_{\bf k}=0$ determine the phase boundary between the CDW and superfluid phases, which is $\mu=\frac{3V}{2}-\sqrt{(\frac{3V}{2})^2-(3t)^2}$ (the maximum value of $|f({\bf k}|)$ is $3t$). We plot the phase boundary from the spin-wave approximation in Fig.\ref{afig8}. It is qualitatively consistent with the exact phase diagram from the QMC method except that the range in the chemical potential is slightly increased. We also determine the phase boundary of the $\rho=\frac{1}{2}$ domain-wall phase on a $W=12$ zigzag ribbon, which greatly shrinks from that of a periodic system.

The Berry curvature associated with each magnon band is given by
\begin{eqnarray}\label{eqb8}
\Omega_{\lambda}({\bf k})=\frac{\partial A_{y}({\bf k})}{\partial k_x}-\frac{\partial A_{x}({\bf k})}{\partial k_y},
\end{eqnarray}
where $A_{i}=-{\rm i}\langle u_{\lambda,{\bf k}}|\frac{\partial}{\partial k_i}|u_{\lambda,{\bf k}}\rangle$ ($i=x,y$) is the Berry potential, and $\lambda=\pm$ denotes the two magnon bands.

\section{5. The domain-wall superfluid}
To demonstrate the domain-wall superfluid clearly, we perform QMC
simulations on a $W=12$ and $L=24$ ribbon. Since there are more
approximately degenerate $\rho=\frac{1}{2}$ phases on wider ribbons, a
small pinning field is used to select a specific
configuration\cite{assaad13}. The
pinning field is a staggered potential with the value $-|\Delta|$
($|\Delta|$) on each occupied (unoccupied) site of the targeted
domain-wall phase. In the following simulations, the strength of the
pinning field is set to $|\Delta|=0.1$, and we focus on the phase with
the domain wall in the middle.

\begin{figure}[htbp]
\centering \includegraphics[width=6.5cm]{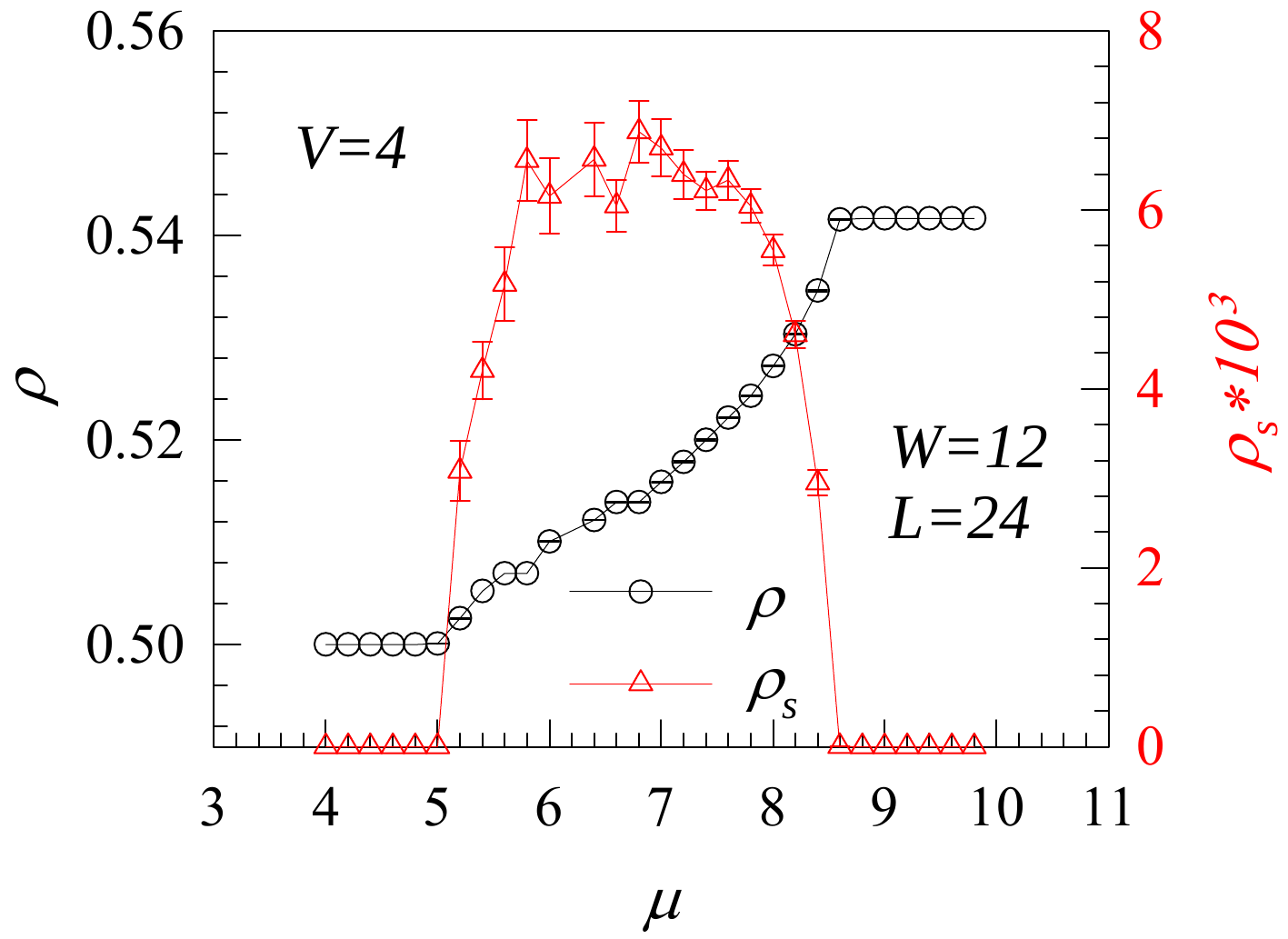} \caption{The average
density and superfluid density as a function of $\mu$ at $V/t = 4$ on a
$W=12$ and $L=24$ ribbon.}
\label{afig9}
\end{figure}

Figure \ref{afig9} shows the average density and superfluid density as a
function of chemical potential at $V/t =4$. As the chemical potential
increases, bosons are continuously added to the $\rho=\frac{1}{2}$
domain-wall phase until the domain wall is full, and the ribbon becomes a
$\rho=\frac{1}{2}+\frac{1}{2W}$ insulator. Between the two insulators,
the superfluid density is nonzero, implying the system is a superfluid.

\begin{figure}[htbp]
\centering \includegraphics[width=7.5cm]{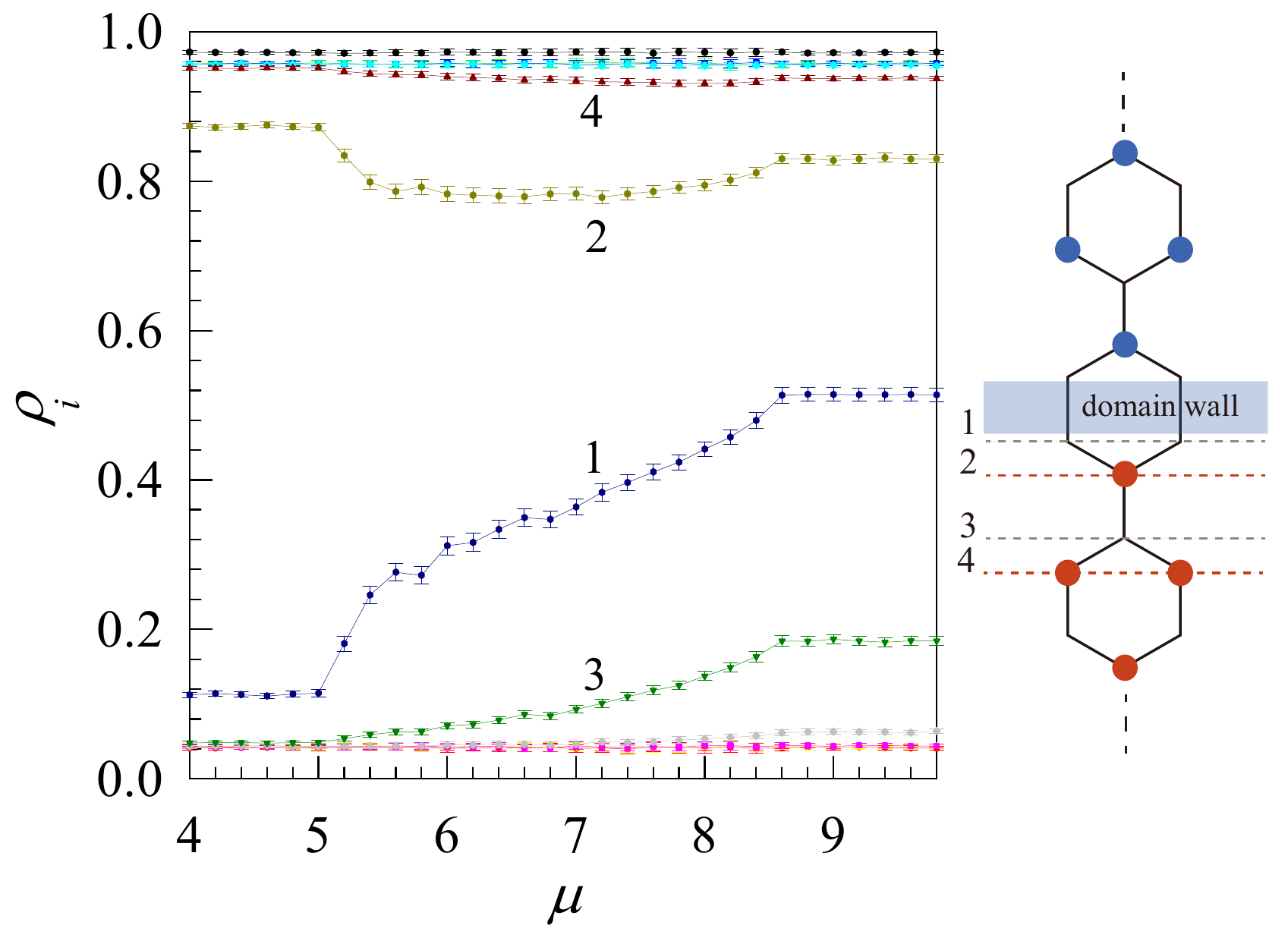} \caption{The local
densities as functions of chemical potential. Due to the translation
symmetry, the sites of each sublattice on the lines parallel to the
domain wall are equivalent (see the right figure). In addition, the system
is symmetric about the domain wall. So we only show the values on
nonequivalent sites, which are from upper(lower)-half unit cell. The
parameters are the same as those in Fig.~\ref{afig9}.}
\label{afig10}
\end{figure}

Next we demonstrate the superfluid is localized near the domain wall,
forming a one-dimensional superfluid channel. Figure \ref{afig10} shows the local
densities as functions of chemical potential. For the
$\rho=\frac{1}{2}$ insulator, the profile of the local density indicates
it is a domain-wall phase with the domain wall in the middle of the
ribbon. From about $\mu=5$, additional bosons are added to the system.
The value on the low-occupation sites of the domain wall increases
significantly. In the $\rho=\frac{1}{2}+\frac{1}{2W}$ insulator, the
site densities
approach  $\rho_i \sim 0.5$, which corresponds to the case the domain
wall is full. The local densities are only slightly affected for
sites more than $2a$ ($a$ the lattice constant) away from the domain
wall. Thus the added bosons mainly reside near the domain wall. It is also
noted that the local densities on the highly-occupied sites near the
domain wall first decrease.  This implies the bosons begin to flow between
high- and low-occupation sites, which is consistent with the
appearance of superfluid. Since the decrease is most significant on the
domain wall, the gapless superfluid is mainly around it.

We also calculate the single-particle correlator. As the distance
$r_{\bot}$ away from the domain wall increases, the decay of the
correlator becomes rapid, suggesting the superfluid density decreases.
For the zigzag chain on the domain wall, the decay of the correlator is
slower than power law. However the decay becomes exponential from
the fourth zigzag chain, where the superfluid begins to vanish. Moreover
the curves remain almost unchanged for the far zigzag chains, manifesting
the uniform bulk CDW order there. Inset shows $\langle
b^{\dagger}_{0}b^{\phantom{\dagger}}_{r}\rangle$ as a function of $r_{\bot}$ (the distance
away from the domain wall) at fixed $r=6$. The data are well fit
using an exponential decay. Thus the superfluid is localized near the
domain wall, and decays exponentially into the bulk.

\begin{figure}[htbp]
\centering \includegraphics[width=7.5cm]{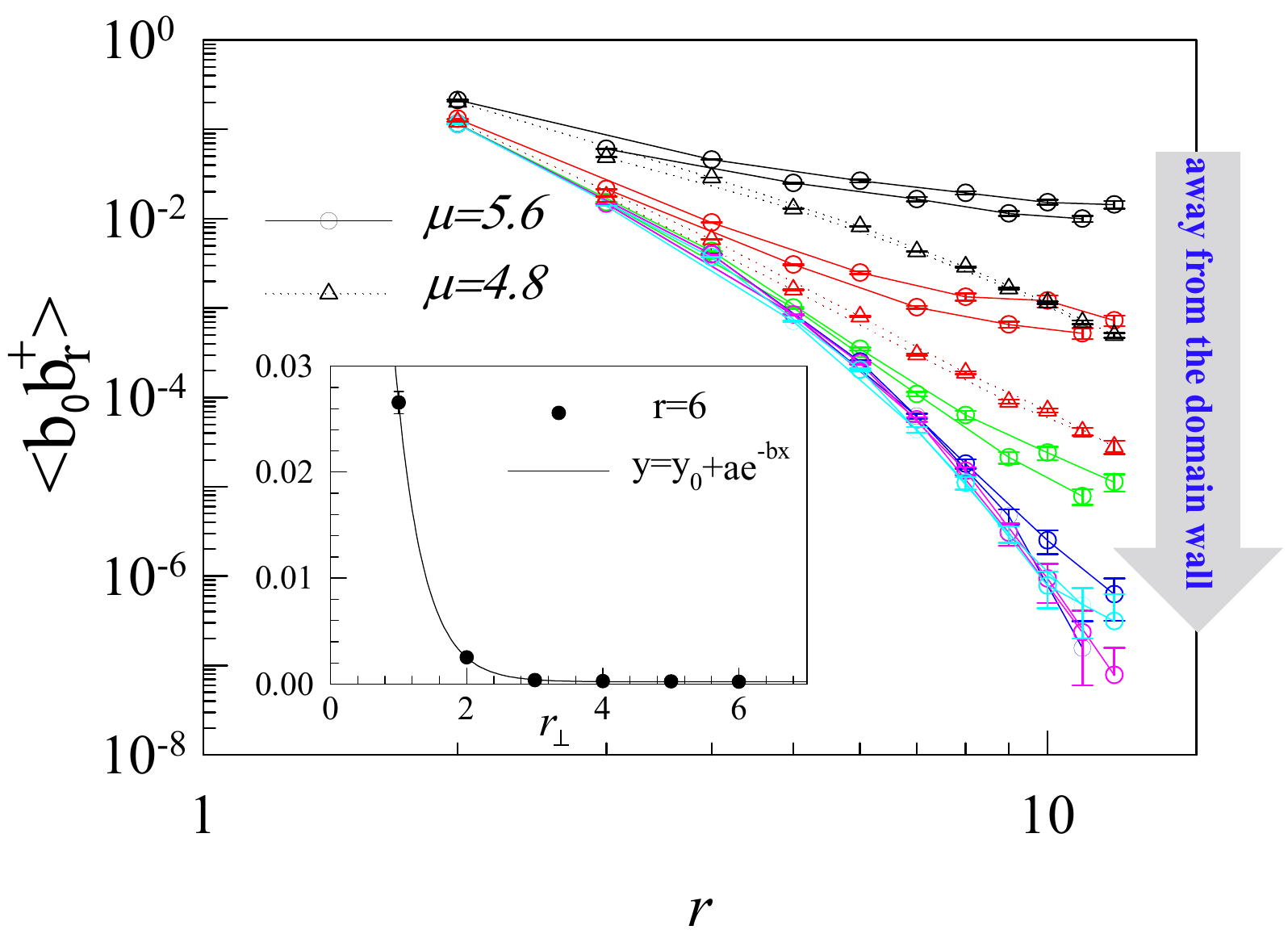} \caption{The
single-particle correlator
$\langle b^{\dagger}_{0}b_{r}\rangle$ as a
function of distance from the reference point. Due to the symmetries,
only nonequivalent zigzag lines are shown. For comparison, we also
   calculate $\langle b^{\dagger}_{0}b_{r}\rangle$ in the
$\rho=\frac{1}{2}$ CDW phase, and show the results along two zigzag
chains closest to the domain wall. Inset shows $\langle
b^{\dagger}_{0}b_{r}\rangle$ as a function of distance $r_{\bot}$ away
from the domain wall at fixed $r=6$. The data are well fit by an
exponential decay.  Here the parameters are the same as those in
Fig.~\ref{afig9}.}
\label{afig11}
\end{figure}

\section{6. The domain wall with turns}
\begin{figure}[htbp]
\centering \includegraphics[width=7cm]{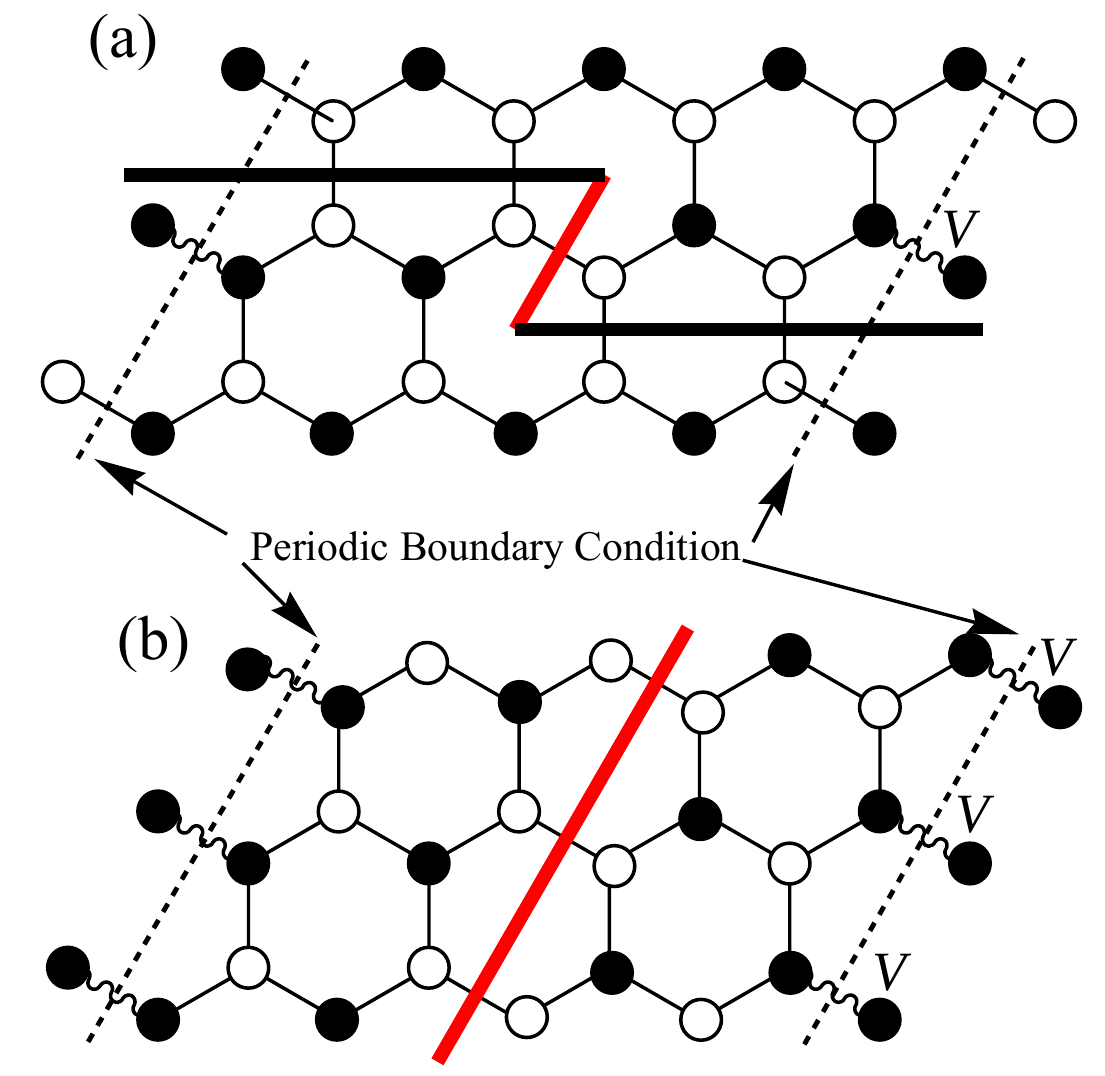} \caption{Schematical demonstrations of: (a) a domain wall with two turns; (b) a domain wall along a direction other than the longer axis of the ribbon. Both configurations are not energetically favored. The two plots are based on a $W=3$ and $L=4$ ribbon. The wavy bonds cross the periodic boundary, each of which connects two occupied sites generating an interacting energy $V$. }
\label{afig12}
\end{figure}

It is interesting to see whether it is possible to have domain walls with turns. The domain wall organized by the NN interactions should be straight and along the ribbon. The open zigzag edges of the ribbons are created by removing a line of vertical bonds. Then both sites connected by such a bond can be occupied. To be at half filling, we have a line of vertical bonds empty in the bulk, which form the "self-organized" domain wall. Once the domain wall has turns, non-vertical bonds crossed should be empty. To maintain the number of bosons at half filling, there must appear adjacent occupied sites connected by periodic boundary condition, which induce interacting energy. As shown in Fig.\ref{afig12}(a), a "Z" shaped domain wall crosses one non-vertical bond, and there appear a pair of adjacent bosons generating an interaction $V$. A domain wall along a direction other than the ribbon is shown in Fig.\ref{afig12}(b). It crosses three non-vertical bonds, and there appear three pairs of adjacent bosons with an interacting energy $3V$. Generally when a domain wall crosses the non-vertical bonds for $n$ times, there appear $n$ pairs of adjacent bosons, and the interacting energy increases by $nV$. So for large interactions, a domain wall with turns is not energetically favored.

\begin{figure}[htbp]
\centering \includegraphics[width=8.5cm]{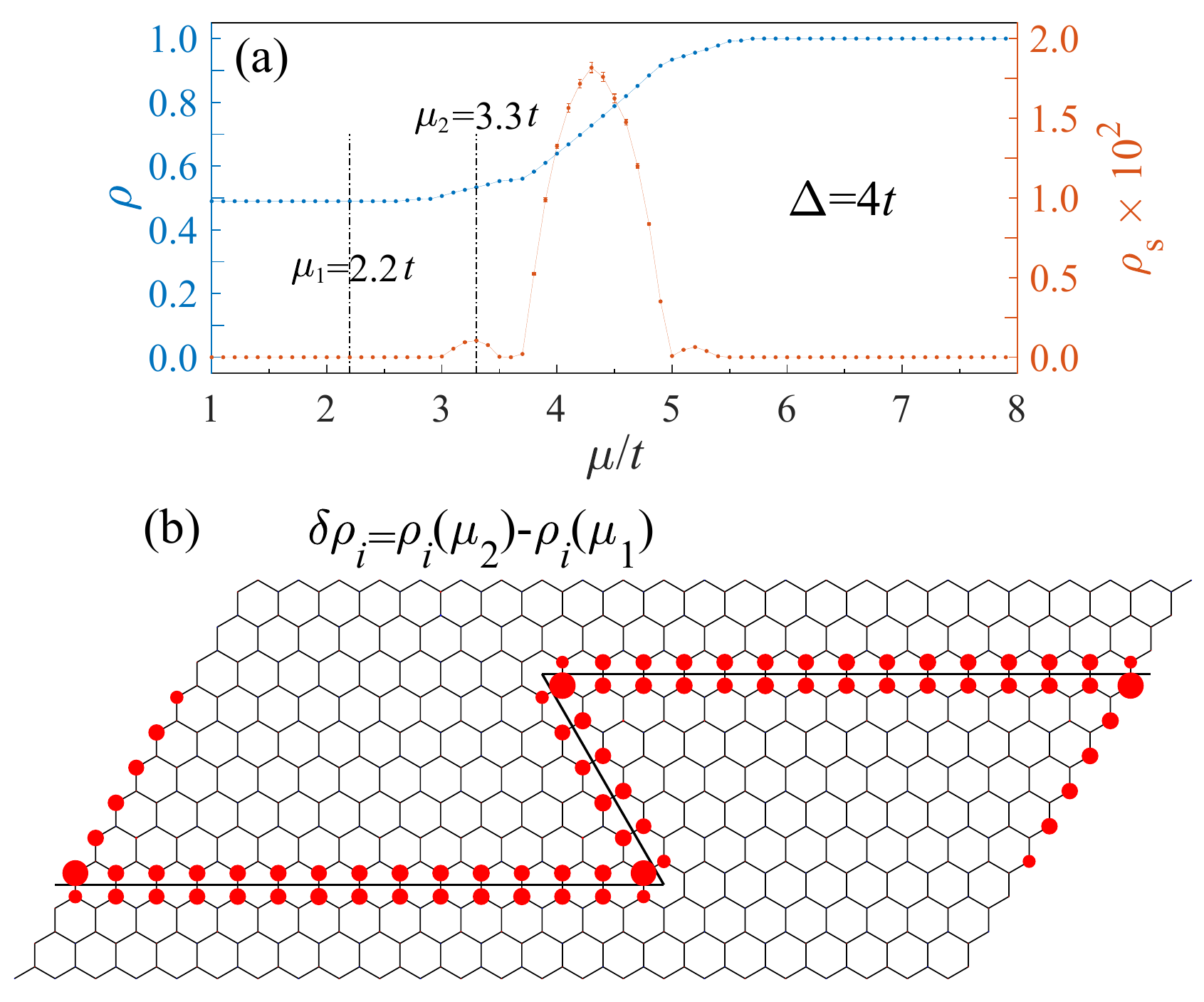} \caption{(a) The average
density and superfluid density as a function of $\mu$ on a ribbon with a "Z" shaped domain wall generated using a staggered potential. (b) The distribution of the bosons added to the $\rho=\frac{1}{2}$ insulator. Ribbon width $W=12$ and length $L=24$. The strength of the staggered potential is $\Delta=4t$.}
\label{afig13}
\end{figure}

Although the domain walls with turns can not be generated by the NN interactions, they can be designed using staggered potentials. We create a "Z" shaped domain-wall using staggered potentials with the strength $|\Delta|=4t$ on a $W=12$ and $L=24$ ribbon, which is periodic along the longer axis. Figure \ref{afig13}(a) plots the average density and the superfluid density as a function of $\mu$. The $\rho=\frac{1}{2}$ plateau with vanishing superfluid corresponds to the domain-wall insulator. As the chemical potential further increases, bosons are added to the $\rho=\frac{1}{2}$ insulator. The superfluid density has finite values, and the system becomes superfluid. We calculate the distribution of the added bosons, $\delta \rho_{i}=\rho_{i}(\mu_2)-\rho_{i}(\mu_2)$, with $\mu_2, \mu_1$ marked in Fig.\ref{afig13}(a). As shown in Fig.\ref{afig13}(b), the added bosons mainly distribute near the $Z$ shaped domain wall, implying the superfluid above the $\rho=\frac{1}{2}$ transporting down such a domain wall. With this method, complex shaped domain wall can be designed, realizing circuits of superfluid.

\end{document}